\documentclass[journal = langd5, manuscript = article]{achemso}
\pdfoutput=1
\usepackage[version=3]{mhchem}
\usepackage{amsmath}
\usepackage{amssymb}
\usepackage{caption}
\usepackage{float}
\usepackage{geometry}
\usepackage{setspace}
\usepackage{xkeyval}
\usepackage{subfigure}
\usepackage{multirow} 
\usepackage{booktabs}
\usepackage[usenames,dvipsnames]{xcolor}
\usepackage{soul}               
\usepackage{color,colordvi}     
\usepackage{times}              
\usepackage{textcomp}           
\SectionNumbersOn
\author{Siddarth A.Vasudevan}
\affiliation{Laboratory for Interfaces, Soft matter and Assembly, Department of Materials, ETH Z{\"u}rich, Vladimir-Prelog-Weg 5, 8093 Z{\"u}rich, Switzerland}
\author{Astrid Rauh}
\affiliation{Physical Chemistry I, University of Bayreuth, Universit{\"a}tsstr. 30, 95440 Bayreuth, Germany}
\alsoaffiliation{Physical Chemistry I, Heinrich-Heine-University, Universit{\"a}tsstr. 1, 40204 D{\"u}sseldorf, Germany}
\author{Martin Kr\"{o}ger}
\affiliation{Polymer Physics, Department of Materials, ETH Z{\"u}rich, Leopold-Ruzicka-Weg 4, 8093 Z{\"u}rich, Switzerland}
\author{Matthias Karg}
\affiliation{Physical Chemistry I, Heinrich-Heine-University, Universit{\"a}tsstr. 1, 40204 D{\"u}sseldorf, Germany}
\author{Lucio Isa}
\affiliation{Laboratory for Interfaces, Soft matter and Assembly, Department of Materials, ETH Z{\"u}rich, Vladimir-Prelog-Weg 5, 8093 Z{\"u}rich, Switzerland}
\email{lucio.isa@mat.ethz.ch}
\title{Dynamics and wetting behavior of soft particles at a fluid-fluid interface\footnote{Supporting information available}}

\begin{document}
	\begin{abstract}
We investigate the conformation, position, and dynamics of core-shell nanoparticles (CSNPs) composed of a silica core encapsulated in a cross-linked  poly-\textit{N}-isopropylacrylamide shell at a water-oil interface for a systematic range of core sizes and shell thicknesses. We first present a free-energy model that we use to predict the CSNP wetting behavior at the interface as a function of its geometrical and compositional properties in the bulk phases, which gives good agreement with our experimental data. Remarkably, upon knowledge of the polymer shell deformability, the equilibrium particle position relative to the interface plane, an often elusive experimental quantity, can be extracted by measuring its radial dimensions after adsorption. For all the systems studied here, the interfacial dimensions are always larger than in bulk and the particle core resides in a configuration wherein it just touches the interface or is fully immersed in water. Moreover, the stretched shell induces a larger viscous drag at the interface, which appears to depend solely on the interfacial dimensions, irrespective of the portion of the CSNP surface exposed to the two fluids. Our findings indicate that tailoring the architecture of CSNPs can be used to control their properties at the interface, as of interest for applications including emulsion stabilization and nanopatterning.
	\end{abstract}

\section{Introduction}
\begin{figure}[htp]
	\centering
	\includegraphics {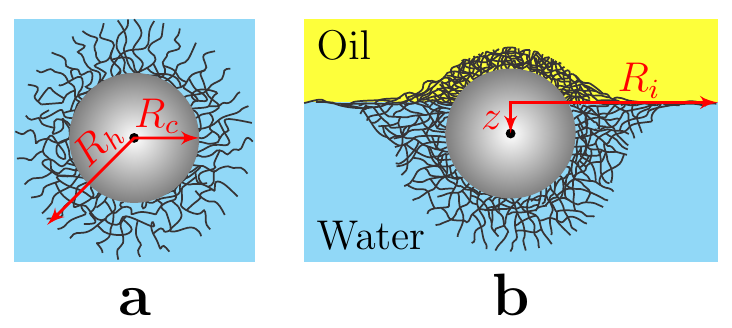}
	\caption{Schematic representation of a CSNP {\bf (a)} in water and {\bf (b)} at the water-oil interface. $R_{c}$ and $R_{h}$ are the core and hydrodynamic radius of an individual CSNP, respectively.  $z$ is the height (negative distance) of the center of the core with respect to the plane of the interface and $R_{i}$ is the radius of the CSNP at the interface.}
	\label{fig: schematic representation of csnp in bulk and water-oil interface}
\end{figure}

Recently, there has been a surge of interest in the study of soft colloidal particle systems, whose chemical and physical properties can be tuned by changing ambient conditions, such as pH, solvent quality, and temperature. Such particles are typically composed of a cross-linked, swollen polymer network and are termed microgels \cite{2017Plamper,2014Yunker}. However, they may also consist of a non-deformable inorganic core encapsulated inside a shell comprising a cross-linked polymer network; we term those here core-shell nanoparticles (CSNPs) (Fig.~\ref{fig: schematic representation of csnp in bulk and water-oil interface}a)\cite{2016Karg}.
 These soft particles, when adsorbed at a fluid interface, undergo a shape transformation, i.e., the polymer network exposed to the interface stretches due to the action of interfacial tension and the particle may also adopt an anisotropic shape perpendicular to the interface plane depending on the difference in the solvent quality of the two fluid phases forming the interface (Fig.~\ref{fig: schematic representation of csnp in bulk and water-oil interface}b) \cite{2011Destribats,2016Rauh,2015Style}. Experimentally, it has been found that the extent of deformation in the plane of the interface depends on the cross-linking density of the polymer network, i.e., its elasticity \cite{2011Destribats}. Moreover, a recent numerical work by Mehrabin et al. showed that, in addition to elasticity, the deformation of a soft particle at a fluid interface is also influenced by its wetting behavior\cite{2016Mehrabin}. In particular, they found that partial wetting of the soft particle leads to small deformations, which can be accurately captured by continuum elastic theory, while large deformations similar to that observed in experiments can be accounted for only if the particle undergoes complete wetting -- wherein continuum elastic theory breaks down and molecular details of the particle architecture, such as the topology of cross-links, become important.

Following deformation, the size acquired by a soft particle at the interface also depends on its position relative to the interface plane, i.e., on whether it protrudes equally into the two fluid phases forming the interface or if it exhibits preferential protrusion into one of the two fluids. Even though theoretical studies typically consider only the symmetrical case \cite{2015Style,2016Mehrabin}, in experiments, preferential protrusion in one fluid, typically water, is most often observed. Currently, there have been only few experimental studies, which have provided qualitative and quantitative measures of microgel position with respect to the interface plane \cite{2012Geisel,2015GeiselTXM,2016Kwok}. Geisel et al. found that, while their microgels were pH-responsive in bulk aqueous dispersions, neither their size nor their protrusion at interface were dependent on  pH \cite{2012Geisel}, emphasizing the dominant role played by interfacial effects. It is important to note that they calculated the protrusion height of microgels indirectly from the interfacial size obtained from freeze-fracture cryo-SEM images under the simplified assumption that the deformed particles took the shape of a spherical cap. Kwok et al. observed similar behavior for their pH-responsive microgels, however in certain cases, they found that large, micron-sized microgels exhibited smaller sizes at the interface when compared to their size in bulk water phase\cite{2016Kwok}. The size of soft particles at a fluid interface is also important for many technological applications. For instance, this parameter has been found to influence the stability of emulsion droplets coated by microgels \cite{2012Richtering} and determines the maximum achievable spacing in soft colloidal lithography templates \cite{2016Rey,2017Scheidegger}. In general, open questions remain on how the bulk size of these soft particle influences their size and position at the interface between two fluids with different solvent qualities. Particularly, for the case of CSNPs, the effect of core size has remained largely unexplored. Moreover, in contrast to the case of  non-deformable colloids\cite{2016Dorr,1995Danov,2006Fischer}, very little is known on the viscous drag experienced by soft particles at fluid interfaces and on how their conformation acquired at the interface couples to their dynamics within the interface plane.

In this manuscript, we investigate the dynamics and wetting behavior of soft CSNPs at a water-oil interface by combining experiments and predictions from a simple Flory-type theoretical model. The experimental study covers both the dynamics and the wetting behavior, and is performed on a well-defined system of CSNPs comprising a silica core of controlled size encapsulated inside a poly-\textit{N}-isopropylacrylamide (PNIPAM) shell of varying thickness. In particular, we study the behavior of CSNPs with two different core sizes and four different shell thicknesses for each core size. The experimental results on the wetting are interpreted within the framework provided by the model, which produces general predictions for a broad range of CSNPs and microgels, also beyond the ones used in our experiments. In Section~\ref{sec:materials and methods}, we describe the CSNP synthesis procedure, and techniques used to characterize their size and dynamics, both in bulk water and at the water-oil interface. In addition, this section presents the key elements of the model, which is then used to estimate the free energy of a CSNP at a liquid-liquid interface as function of its position with respect to the interface plane. In Section~\ref{subsec:Wetting behavior}, we present the model predictions for the interfacial equilibrium position and size for a broad range of CSNPs with variable core sizes and shell thicknesses ranging from 50 nm to 500 nm, followed by an analytical calculation for the equilibrium position of a microgel. The predictions of the model are then compared to the experimental data. The analysis of the wetting behavior is followed by the experimental results on the bulk and interfacial dynamics of our CSNPs in Section~\ref{subsec:dynamics}. In Section~\ref{sec:Conclusions} we summarize our main findings and their implications for future studies.

\section{Materials and Methods}
\label{sec:materials and methods}

\subsection{Materials}
Tetraethylorthosilicate (TEOS; Sigma-Aldrich; 98\%), ammonium hydroxide solution (NH$_{3}$ (aq.); Sigma-Aldrich; 30-33\%), rhodamine b isothiocyanate (RITC; Sigma-Aldrich; mixed isomers), (3-aminopropyl)trimethoxysilane (APS; Sigma-Aldrich; 97\%), 3-(trimethoxysilyl)propyl methacrylate (MPS; Sigma-Aldrich; 98\%), ethanol (EtOH; Sigma-Aldrich; $\geq$99.8\%), sodium dodecyl sulfate (SDS; Merck; Ph. Eur.), \textit{N}-isopropylacrylamide (NIPAM; Sigma-Aldrich; 97\%), \textit{N},\textit{N}'-methyl-enebisacrylamide (BIS; Fluka; $\geq$98\%), hexane (Sigma-Aldrich; 99\%) and potassium peroxodisulfate (PPS; Fluka; $\geq$99\%) were used as received. Water was purified using a Milli-Q system (18 M$\Omega$ cm). Hexadecane (Sigma-Aldrich; 99\%) was purified to remove surface-active contaminants by passing it through a column containing both alumina (MP Biomedicals; MP EcoChrome Alumina B) and silica (Fluka; 60 \AA{} pores, 70-230 mesh).

\subsection{Methods}
\subsubsection{Synthesis and functionalization of the silica particles}

We synthesized three different batches of fluorescent silica particles, which are denoted as Core1 (C1), Core2 (C2), and Core3 (C3) in Table~\ref{tbl: Tabulation volumes of materials used in silica particle synthesis}, following a recently published protocol\cite{2016Rauh}. Prior to the St\"{o}ber synthesis of the silica particles, we functionalized the RITC dye. A ten-fold excess of APS was added dropwise to a 10 mM ethanolic RITC solution to ensure covalent binding to the dye molecule. The mixture was then stirred in the dark for at least 2 h. 333 \textmu{}L of this dye solution was diluted with ethanol in a ratio of 1:5 before adding it during the silica particle synthesis. Two different solutions, detailed in Table~\ref{tbl: Tabulation volumes of materials used in silica particle synthesis}, were prepared simultaneously in order to synthesize the silica colloids. Solution-1 was mixed in a three-neck round-bottom flask before heating to 50 \textdegree{}C, while solution-2 was prepared by heating TEOS and ethanol to 50 \textdegree{}C and equilibrating for 20 min. After preparation, solution-2 was quickly added to solution-1. As silica seeds formed, the reaction mixture turned turbid; at this moment, the addition of the dye solution was started. The reaction was allowed to proceed for 24 h, after which the mixture was cooled to room temperature and purified twice by centrifugation and subsequent redispersion in ethanol. The functionalization of silica particles using MPS was performed as described in reference\cite{2016Rauh}. The final concentration of the different silica seed stock dispersions and the radius of the silica cores obtained from SEM images are also reported in Table~\ref{tbl: Tabulation volumes of materials used in silica particle synthesis}. Since the core radius of C2 is nearly the same as that of C3, we will refer to both batches as particles with core radius of 176 nm in Section~\ref{sec:results and discussion}.
\begin{table}[htp]
	\small
	\caption{Volumes of EtOH, NH$_3$ (aq., 30-33\%), H$_2$O, and TEOS used to synthesize the silica particles and the concentration of the seed stock solutions.}
	\label{tbl: Tabulation volumes of materials used in silica particle synthesis}
	\begin{tabular*}{0.96\textwidth}{@{\extracolsep{\fill}}cc|ccc|cc|c}
		\hline\hline
		\multirow{3}{*} {Silica colloids} & Core & \multicolumn{3}{c|}{Solution-1} & \multicolumn{2}{c|}{Solution-2} & Conc. of \\
		& size & EtOH & NH$_3$, aq. & H$_2$O & EtOH & TEOS & seed stock soln.\\
		& [nm] & [mL] & [mL] & [mL] & [mL] & [mL] & [\textmu{}M]\\
		\hline
		Core1 (C1) & $63\pm4$ & 125 & 10 & - & 20 & 5 & 0.0197\\
		Core2 (C2) & $174\pm9$ & 37.6 & 10.7 & 18.3 & 26.8 & 6.7 & 0.0290\\
		Core3 (C3) & $176\pm8$ & 56.4 & 27.45 & 16.05 & 40.2 & 10.05 & 0.0184\\
		\hline\hline
	\end{tabular*}
\end{table}

\subsubsection{Synthesis of SiO$_2$-PNIPAM particles}
\label{subsubsec:standard csnp synthesis procedure}
\textit{Standard procedure}\newline
Encapsulation of the functionalized silica particles in a cross-linked PNIPAM shell was performed by using free radical seeded precipitation polymerization; details on the amounts of different materials used for each CSNP can be found in Table~2 and 3. Particles are named according to their core size and shell thickness, e.g., C1S1 corresponds to CSNPs with core C1 and shell thickness S1. The synthesis was conducted in a three-neck round-bottom flask equipped with a reflux condenser and magnetic stirrer. Specific amounts of NIPAM, BIS, and 0.2 mM SDS were dissolved in water while stirring. The solution was heated to 70 \textdegree{}C and purged with nitrogen to remove oxygen. The reaction mixture was then allowed to equilibrate for 20 min. Next, a specific quantity of the silica seed stock dispersion was added. After further equilibration for 15 min, PPS dissolved in 1 mL of water was added quickly. After the reaction proceeded for 2 h, the mixture was cooled to room temperature and purified thrice by centrifugation and subsequent redispersion of the sediment in water.
\begin{table}[htp]
	\small
	\caption{Quantities of NIPAM, BIS, PPS, H$_2$O, and  silica stock dispersion (SiO$_{2}$) used to prepare C1 and C2 CSNPs.}
	\label{tbl:csnp prep table1}
	\begin{tabular*}{0.48\textwidth}{@{\extracolsep{\fill}}c|ccc|cc}
		\hline\hline
		\multirow{3}{*}{CSNP} & \multicolumn{3}{c|}{Mass} & \multicolumn{2}{c}{Volume}\\
		 & NIPAM & BIS & PPS & H$_2$O & SiO$_{2}$\\
		& [mg] & [mg] & [mg] & [mL] & [\textmu{}L] \\
		\hline
		C1S1 & 91 & 6.2 & 4 & 40 & 1125\\
		C1S2 & 136 & 9.3 & 4 & 40 & 1125\\
		C1S3 & 113 & 7.7 & 2 & 20 & 438\\
		C1S4 & 113 & 7.7 & 2 & 20 & 250\\
		\hline
		C2S1 & 68 & 4.6 & 2 & 20 & 323\\
		C2S2 & 113 & 7.7 & 2 & 20 & 323\\
		\hline\hline
	\end{tabular*}
\end{table}
\newline\newline
\textit{Semi-batch seeded precipitation polymerization}\newline
SiO$_2$-PNIPAM particles with thicker polymer shells were synthesized using a semi-batch method to avoid agglomeration and the formation of purely organic microgel particles. Both problems can occur in a one-step procedure due to the very high monomer concentrations that are necessary to achieve thick polymer shells. Initially, the same steps described in the standard procedure (see previous paragraph) were performed for the basis reaction. 45 min after the initiation, SDS, NIPAM, BIS, and PPS were added sequentially. At first, a respective amount of SDS was dissolved in 2 mL of water. This SDS solution was added dropwise to the reaction mixture to stabilize the particles during the monomer additions. Then, BIS was dissolved in 2 mL of water and NIPAM was dissolved in 4 mL of water. A syringe pump was used to add the NIPAM solution within 30 min. 1 mL of the prepared BIS solution was added to the reaction mixture when the NIPAM addition started. Subsequently, PPS dissolved in 1 mL of water was used for initiation. Afterwards, the residual milliliter of BIS solution was added dropwise. 45 min after the last initiation the next monomer addition was performed following the same procedure. The respective quantities of used chemicals for the basis reaction as well as for the addition steps are summarized in Table 3. Purification of these particles was performed in the same manner as for the CSNPs prepared using the standard procedure, i.e., via three runs of centrifugation and redispersion in water.
\begin{table}
	\small
	\caption{Quantities of NIPAM, BIS, SDS, PPS, H$_2$O, and silica stock dispersion (SiO$_{2}$) for preparation of C3S1 and C3S2 CSNPs.}
	\label{tbl:csnp prep table2}
	\begin{tabular*}{0.96\textwidth}{@{\extracolsep{\fill}}cc|cccc|cc}
		\hline\hline
		\multirow{3}{*}{CSNP} & \multirow{3}{*}{Synthesis step} & \multicolumn{4}{c|}{Mass} & \multicolumn{2}{c}{Volume}\\
		&  & NIPAM & BIS & SDS & PPS & H$_2$O & SiO$_{2}$\\
		&  & [mg] & [mg] & [mg] & [mg] & [mL] & [\textmu{}L] \\
		\hline
		\multirow{2}{*}{C3S1} & Basis reaction & 113 & 7.7 & 1.2 & 2 & 20 & 1018\\
		& Addition 1 & 113 & 7.7 & 1.2 & 2 & $+$8 & -\\
		\hline
		\multirow{5}{*}{C3S2} & Basis reaction & 113 & 7.7 & 1.2 & 2 & 20 & 1018\\
		& Addition 1 & 170 & 11.6 & 1.7 & 3 & $+$8 & -\\
		& Addition 2 & 226 & 15.4 & 2.3 & 4 & $+$8 & -\\
		& Addition 3 & 283 & 19.3 & 2.9 & 5 & $+$8 & -\\
		& Addition 4 & 339 & 23.1 & 3.5 & 6 & $+$8 & -\\
		\hline\hline
	\end{tabular*}
\end{table}
\subsubsection{Bulk sizing and dynamics}
\label{subsubsec: dynamics and sizing in the bulk}
The hydrodynamic radius $R_h$ of the CSNPs in water was obtained by dynamic light scattering using a Zetasizer Nano ZS device equipped with a 633 nm laser. The values of $R_h$ were recorded as a function of temperature ranging from 20 \textdegree{}C to 50 \textdegree{}C in steps of 3 \textdegree{}C. Five measurements were performed at each temperature, from which the average value of $R_h$ was obtained. For each measurement at a given temperature the standard deviation in the hydrodynamic radius, $\Delta R_h$, was obtained from the polydispersity index (PdI) using the following relationship: $\Delta R_{h} = R_{h}\sqrt{\mathrm{PdI}}$. The best estimate for $\Delta R_h$ from the five measurements was obtained by using the quadrature rule for error propagation \cite{1982Taylor}.

We used two different variants of differential dynamic microscopy (DDM) to characterize the dynamics of CSNPs in bulk water. To this end, we performed brightfield-DDM (b-DDM) and darkfield-DDM (d-DDM)\cite{2008Cerbino,2016Bayles}, respectively, on CSNPs with core radius of 63 nm (C1) and 176 nm (C2/C3). The samples for DDM measurements were prepared by filling dilute aqueous CSNP suspensions into rectangular capillary tubes with dimensions of 100 \textmu{}m $\times$ 50 mm $\times$ 2 mm (thickness, length, and width, respectively). We briefly describe the DDM technique here.

In a typical DDM measurement and analysis procedure, we obtained a time sequence of real-space images of the suspension. Subsequently, these images were Fourier-transformed to obtain the wavenumber-dependent dynamics. For each value of the wavenumber $q$, time and azimuthal-averaged operations were performed to obtain the intermediate scattering function (I.S.F.), which is dependent on both $q$ and the lag-time $\Delta t$. For the b-DDM measurements, 3000 images were acquired and 1000 time-average operations were performed for each value of $\Delta t$, while for the d-DDM measurements 10000 images were acquired and 6000 time-average operations were performed for each value of $\Delta t$. All the image sequences were acquired at a rate of 100 frames per second with each image containing 256$\times$256 pixels. The effective pixel sizes for b-DDM and d-DDM measurements were 0.10 \textmu{}m (63$\times$ objective) and 0.16 \textmu{}m (40$\times$ objective), respectively.

\subsubsection{Depositions and AFM imaging}
\label{subsubsec: depositions and afm imaging}
To determine the size of the C1 CSNPs at the water-hexadecane interface, we used an ex-situ method wherein the particles were deposited from the interface onto a silicon wafer. The deposition was achieved by using the following procedure. A holder containing a clean hydrophilic silicon wafer at a small tilt angle was placed into a teflon container. The holder itself was connected to a linear-motion driver. Then we filled the container with Milli-Q water until the silicon wafer was completely immersed. Hexane was gently added onto the surface of water to create a water-hexane interface. Subsequently, using a syringe, we injected a 1:1 mixture of isopropanol and an aqueous suspension of CSNPs at the interface. The system was allowed to equilibrate for few minutes before the linear-motion driver slowly lifted the wafer upwards to collect the particles from the interface. The CSNPs deposited onto the silicon wafer were imaged by using an AFM (Bruker Icon Dimension). We acquired 512 $\times$ 512 pixels$^2$ images with dimensions of 6 $\times$ 6 \textmu{}m$^2$ at a scan rate of 0.5 Hz using a micro-cantilever (Olympus) with resonance frequency around 300 kHz and a spring constant of 26 Nm$^{-1}$. The radius of the particles at the interfacial plane, denoted as $R_i$, was measured from the AFM phase images using Fiji.

\subsubsection{Sizing and Dynamics at the interface}
\label{subsubsec: dynamics and sizing at the interface}
A custom-made cell was designed to prepare samples for measuring the size of the C2/C3 CSNPs and the dynamics of both C1 and C2/C3 CSNPs at the water-hexadecane interface. The cell was fabricated by gluing a cover glass (0.08 mm to 0.12 mm, Thermo Scientific) with a 10 mm diameter hole (cut by using a Hunst laser cutter provided with a $\mathrm{CO_{2}}$ laser of 10600 nm wavelength) and an aluminium ring with an inner diameter of 20 mm onto a cover glass (0.08 mm to 0.12 mm, Thermo Scientific) with 40 mm diameter. The small inner cavity of the cell was filled with either a very dilute (0.1 wt\%) or a relatively concentrated (1 wt\% to 3 wt\%) aqueous CSNP suspension, depending on whether dynamics or sizing measurements were being performed at the interface. The outer cavity delimited by the aluminium ring was filled with hexadecane and sealed on top with a cover glass to avoid convection. Designing the cell in this manner allowed us to access the interface using a 63$\times$ water immersion objective (C-Apochromat, numerical aperture of 1.2) with a free working distance of 0.28 mm. The particles were imaged at the interface by using a Zeiss Spinning Disk confocal microscope equipped with a 561 nm diode laser (200 mW) and an EM-CCD camera (Photometrics Evolve 512). The particle positions were located by post-processing the images using the MATLAB$^\text{\textregistered}$ version of the particle tracking algorithms originally developed by Crocker and Grier \cite{1996Crocker}. The C2/C3 CSNP size at the water-hexadecane interface was determined by measuring the nearest-neighbor distance from a densely aggregated network of particles. The histogram of the nearest-neighbor distances was fitted with a normal distribution to obtain the mean and the standard deviation in the particle diameter at the interface.

To characterize the dynamics of our C1 and C2/C3 CSNPs at the interface, we used a covariance-based estimator during the calculation of their diffusion coefficient. This method was adopted because it provides an optimal way to determine a diffusion coefficient from short particle trajectories, avoiding issues arising from photobleaching. The diffusion coefficient $D$ is estimated from a set of individual particle trajectories using the following relationship:
\begin{eqnarray}
D = \frac{\langle(\Delta x_{i})^2 + (\Delta y_{i})^2\rangle}{4\Delta t} + \frac{\langle\Delta x_{i}\Delta x_{i+1}\rangle}{2\Delta t} + \frac{\langle \Delta y_{i}\Delta y_{i+1}\rangle}{2\Delta t}
\end{eqnarray}
where $\Delta x_i$ and $\Delta y_i$ denote the interfacial displacements of a given particle between frames $i$ and $i-1$, and $\Delta t$ is the lag-time between two successive frames. The $\langle...\rangle$ denotes an average for all particles over the time-series.

\subsubsection{Free energy model of a CSNP at a fluid-fluid interface}
\label{subsec:Modeling}
Here we develop a schematic model to identify the equilibrium position of our deformable CSNPs at a fluid-fluid interface. While the model is more general, we will denote one of the fluids as water and the other one as oil. Water is the better solvent for our CSNP. The model differs from the one in our previous work\cite{2011Isa2,2014Zell} as it considers a cross-linked polymer shell instead of a brush of linear polymer chains grafted onto the nanoparticle core. We consider a CSNP characterized by its core radius $R_c$ and its hydrodynamic (core $+$ shell) radius $R_h$ in  water (Fig.~\ref{fig: schematic representation of csnp in bulk and water-oil interface}a). $R_h$ is for us an experimental input and is not calculated based on the microscopic properties of the shell, e.g. as in \cite{Halperin2011}. The CSNP resides in the vicinity of the water-oil interface (Fig.~\ref{fig: schematic representation of csnp in bulk and water-oil interface}b) with its core center located at a position $z\le 0$ within the water phase; the interface plane defines $z=0$. We consider four contributions to the total free energy $F_\textrm{total}$ of a single CSNP at position $z$ with respect to the interface,
\begin{equation}\label{Eq:Total free energy at interface}
F_\textrm{total}(z) =  F_{w}(z) + F_{o}(z) + F_{e,i}(z) + F_{\gamma}(z).
\end{equation}
There is the free energy of the portion of the shell exposed to the oil phase $F_o$, the free energy of the portion of the shell exposed to the water phase $F_w$, the elastic energy due to the stretching of the shell at the interface $F_{e,i}$ and the free-energy gain $F_\gamma$ obtained by removing the area of the interface occupied by the CSNP. The equilibrium position $z_\textrm{eq}$ will be obtained by minimizing $F_\textrm{total}$ with respect to $z$. To this end, it is convenient to introduce $z$-independent specific free energies $f$ and volumetric free energy densities.

Starting with the first three contributions in eq~\ref{Eq:Total free energy at interface}, we rewrite them more conveniently in terms of densities
\begin{eqnarray}\label{Eq:free energy components}
F_{w}(z) &=& \frac{V_{w}(z)}{\nu_{a}} f_{w}, \nonumber\\
F_{o}(z) &=& \frac{V_{o}(z)}{\nu_{a}} f_{o},\ \textrm{and} \\
F_{e,i}(z) &=& \frac{V_{i}(z)}{\nu_{a}} f_{e,i}. \nonumber
\end{eqnarray}
Here, $V_w$ and $V_o$ denote the $z$-dependent volumes of the polymer shell in water and oil, respectively, while $V_i$ is the volume occupied by the stretched shell at the interface; $\nu_{a}$ is the volume of a single polymer repeat unit. The net contribution of each of the free energy components to the total free energy is dependent on the position of the core with respect to the interface, either via the volume of the shell exposed to each phase or the radius of the CSNP at the interfacial plane. The ratios $f_{w}/\nu_{a}$, $f_{o}/\nu_{a}$, and $f_{e,i}/\nu_{a}$ are volumetric energy densities for $F_w$, $F_o$, and $F_{e,i}$ respectively. Both $f_{w}$ and $f_{o}$ represent specific free energies of polymers residing in a polymer gel, where the solvent is either water or oil. They can thus be modeled by the Flory-Huggins theory \cite{1996Doi} of polymer gels, and are both a sum of elastic $f_e$ and mixing $f_\textrm{mix}$ parts. The mixing part contains Flory's solvent quality parameter $\chi$, which is different for water and oil, i.e. $\chi\ll 0.5$ for good solvents and $\chi\gg0.5$ for poor solvents. Specifically,
\begin{align}
f_{e} &= \frac{3}{2} k_\textrm{B}T\left(\frac{\phi_{0}}{N}\right)\left[\left(\frac{\phi_{0}}{\phi}\right)^{2/3} + \left(\frac{\phi}{\phi_{0}}\right)^{2/3}\right]\label{Eq:elastic free energy}\\
f_\textrm{mix} &=  k_\textrm{B}T\left(\frac{\phi_{0}}{\phi}\right)\left[\left(1-\phi\right)\log \left(1-\phi\right)+\chi \phi\left(1-\phi\right)\right],
\label{Eq:mixing free energy}
\end{align}
where $\phi = (\phi_{w},\phi_{o})$ denotes the volume fraction of the shell in the two different solvents, $\chi=(\chi_w,\chi_o)$ defines the solvent qualities of the two fluids for the polymer gel,
$N$ is the number of polymer repeat units present in a polymer chain segment between two crosslinks of the polymer shell network, and $\phi_0$ is an overall volume fraction of the shell, coinciding with the elastically preferred volume fraction in the absence of mixing terms, in accordance with eq~\ref{Eq:elastic free energy}. The number of partial chains in the gel in contact with water, $V_w(z)\phi_0/N\nu_a$, gives rise to the prefactor in eq\ \ref{Eq:elastic free energy}. $\phi_{w}$ and $\phi_{o}$ are determined separately by minimizing the free energy density $f_{e}+f_\textrm{mix}$ for given values of $\chi$, $N$, $T$, and $\phi_0$, and they do not depend on $z$.

\begin{figure}[htp]
	\centering
	\includegraphics[scale=1]{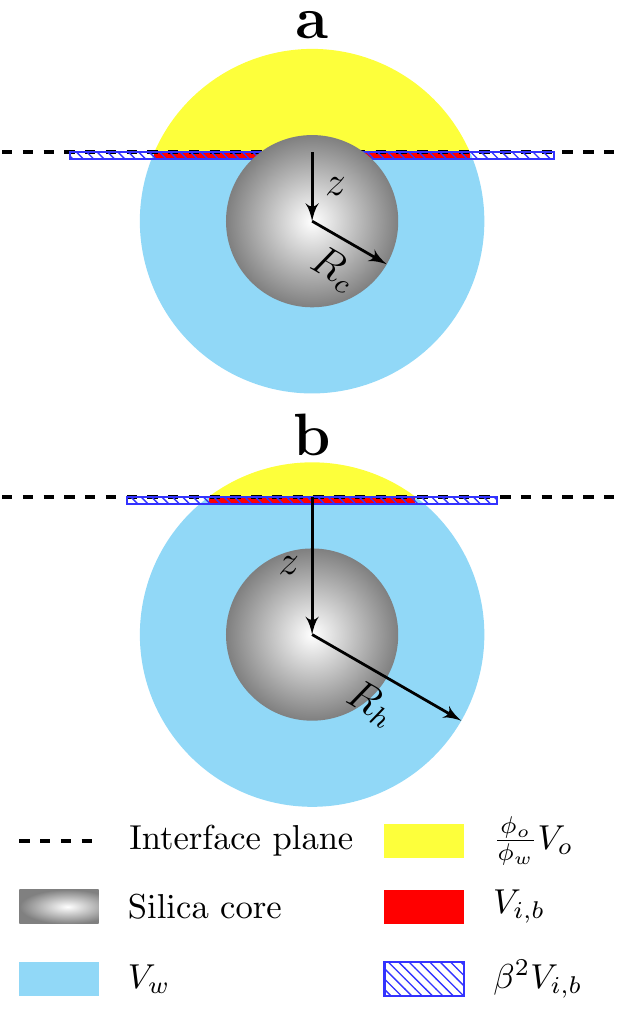}
	\caption{Schematic representation of a CSNP at a water-oil interface, for \textbf{(a)} the core sitting at the interface ($z\in[-R_c,0]$) and \textbf{(b)} the core sitting completely in the water phase ($z\in[-R_h,-R_c]$), with some amount of the shell adsorbed at the interface. The yellow region corresponds to the volume of the shell that was available in the oil phase, i.e., $\frac{\phi_o}{\phi_w}V_{o}$, \emph{before} it underwent deswelling to occupy a volume of $V_{o}$. The red region corresponds to the shell volume shell available at the interface, i.e., $V_{i,b}$, before it undergoes stretching, while the shaded blue region corresponds to the volume after stretching, i.e., $\beta^2V_{i,b}$.}
	\label{Fig:schematic representation of csnp used for modeling}
\end{figure}

The fourth term in eq~\ref{Eq:Total free energy at interface} is due to surface tension, and is thus proportional to the interfacial area occupied by the CSNP,
\begin{equation}
F_{\gamma}(z) = -\pi R_{i}^2(z)\gamma. \label{Eq:free energy from area removal}
\end{equation}
with an interfacial tension $\gamma$ of the oil-water interface, and an effective radius $R_{i}$ of the CNSP within the interface plane. Similar to the quantities defined above, $R_i$ is generally a function of $z$, and can be expressed through the radius of the spherical core, $R_c$, and the hydrodynamic radius $R_h$ of the CNSP dispersed in pure water. One has, according to Figs.\ \ref{Fig:schematic representation of csnp used for modeling}a and \ref{Fig:schematic representation of csnp used for modeling}b,
\begin{equation}\label{Eq:radius of particle at interface}
R_{i}(z) =
\begin{cases}
\sqrt{R_{c}^2-z^2}+\beta\left(\sqrt{R_{h}^2-z^2}-\sqrt{R_{c}^2-z^2}\right), & z\in[-R_c,0]\\
\beta\sqrt{R_{h}^2-z^2}, & z\in[-R_h,-R_c]\\
0, &z\le -R_h
\end{cases}
\end{equation}
where the third case corresponds to a particle fully dispersed in water and fully detached from the interface. In eq.~\ref{Eq:radius of particle at interface}, $\beta$ defines a stretching factor of the polymer shell after adsorption at the interface. While one could expect the polymer shell present at different distances from the contact line to deform to different extents, we assume here that the total stretching can be captured by a single-valued parameter $\beta$. According to our definition, $z$ is zero when the center of the core is located within the plane of the interface and is negative when the core sits preferentially in the water phase. The situation where $z$ is positive -- the center of the core is located in the oil phase -- is not considered here (but trivially reached upon exchanging 'o' with 'w') because this is an unfavorable state for the particle due to the poor solvency of the polymer shell in the oil phase. As seen from eq~\ref{Eq:radius of particle at interface}, for the case of $z\in[-R_c,0]$ (Fig.~\ref{Fig:schematic representation of csnp used for modeling}a), $R_{i}$ has two contributions: a non-deformable component coming from the core $\sqrt{R_{c}^2-z^2}$ and a deformable component corresponding to the length of the shell available at the interface, $\sqrt{R_{h}^2-z^2}-\sqrt{R_{c}^2-z^2}$, which is extended by a yet unknown amount $\beta>1$ due to interfacial tension (see Fig.~\ref{Fig:schematic representation of csnp used for modeling}a). For the simpler case of the core sitting completely in the water phase, i.e., $z\in[-R_h,-R_c]$ (Fig.~\ref{Fig:schematic representation of csnp used for modeling}b), $R_{i}$ has instead only the shell contribution.

The elastic energy penalty coming from the stretching of the shell at the interface, $f_{e,i}V_{i}(z)$, is estimated from the elastic free energy in eq~\ref{Eq:elastic free energy} by using $\phi = \phi_{i}$, where $\phi_{i}$ is the volume fraction of the shell after it stretches at the interface. The required $\phi_{i}$ can be determined from eq~\ref{Eq:phi interface equation}, because the total number of polymer units available at the interface can be assumed conserved during stretching, i.e. $\phi_i V_i=\phi_w V_{i,b}$. More precisely,
\begin{equation}\label{Eq:phi interface equation}
\phi_{i}(z) =
\begin{cases}
\phi_{w}\frac{V_{i,b}(z)}{V_{i}(z)}, \quad &  z\in[-R_h,0]\\
0, &z \leq -R_{h}
\end{cases}
\end{equation}
To make use of eq\ \ref{Eq:phi interface equation}, the volume $V_{i,b}(z)$ of the shell available at the interface before stretching is approximated as an annular disc with thickness $\delta$ (red region in Fig.~\ref{Fig:schematic representation of csnp used for modeling}a) for $z\in[-R_c,0]$ and simply as a disc (red region in Fig.~\ref{Fig:schematic representation of csnp used for modeling}b) of thickness $\delta$ for $z\in[-R_h,-R_c]$, where $\delta$ is essentially the thickness of the interface. The resulting geometric relationships are
\begin{equation}\label{Eq:volume of shell at interface before stretching}
V_{i,b}(z)=
\begin{cases}
\pi \delta \left(R_{h}^2-R_{c}^2\right), &z\in[-R_c,0]\\
\pi \delta\left(R_{h}^2-z^2\right), &z\in[-R_h,-R_c]\\
0, & z \leq -R_{h}
\end{cases}
\end{equation}
so that the volume $V_{i,b}(z)$ varies monotonically with the insertion depth $-z$. The stretching of the shell at the interface only results in an increase in the available volume $V_{i,b}$ by an extent $\beta^2$, implying
\begin{equation}
 V_i(z)=\beta^2 V_{i,b}(z).\label{Eq:volume of shell at interface after stretching}
\end{equation}
Equation~\ref{Eq:phi interface equation} can thus be simplified further by using the latter relationship and eq~\ref{Eq:volume of shell at interface before stretching} to get
\begin{equation}\label{Eq:phi interface equation2}
\phi_{i} =
\begin{cases}
\phi_w/\beta^2, &z\in[-R_h,0]\\
0, & z < -R_{h}
\end{cases}
\end{equation}
Equation~\ref{Eq:phi interface equation2} shows that $\phi_{i}$ is a constant according to eq~\ref{Eq:phi interface equation}, as long as the CSNP remains in contact with the interface. Hence, $f_{e,i}$ is also constant. The $z$-dependent volumes $V_w$ and $\frac{\phi_o}{\phi_w}V_o$ of the shell, indicated by blue and yellow regions in Fig.~\ref{Fig:schematic representation of csnp used for modeling}, can be determined from straightforward geometrical considerations to read
\begin{eqnarray}\label{Eq:volume of shell in water phase}
 V_w(z) &=& \frac{4\pi}{3}(R_h^3-R_c^3) - V_{i,b}(z) - \frac{\phi_o}{\phi_w} V_o(z), \qquad z\le 0.
\end{eqnarray}

The first term is the shell volume of the CSNP fully dispersed in water and the second term, $V_{i,b}$, is the shell volume available at the interface before stretching (eq~\ref{Eq:volume of shell at interface before stretching}). The last term, $\frac{\phi_{o}}{\phi_{w}}V_{o}$, is the shell volume in the oil phase region before deswelling to $V_o(z)$, and is given by
\begin{equation}\label{Eq:volume of shell in oil phase}
\frac{\phi_o}{\phi_w}V_{o}(z) =
\begin{cases}
\frac{\pi}{3}\left(R_{h}+z\right)^2\left(2R_{h}-z\right) - \frac{\pi}{3}\left(R_{c}+z\right)^2\left(2R_{c}-z\right), &z\in[-R_c,0]\\
\frac{\pi}{3}\left(R_{h}+z\right)^2\left(2R_{h}-z\right), &z\in[-R_h,-R_c]\\
0, &z \leq -R_{h}
\end{cases}
\end{equation}

With $V_w(z)$, $V_o(z)$, $V_i(z)$ at hand, with the expression for the specific free energies $f_{w,o}$ given by $f_e(\phi)+f_\textrm{mix}(\phi)$ with $\phi$ obtained as described above, and with the expression for $R_i(z)$ that determines $F_\gamma(z)$, we can directly calculate $z_{\textrm{eq}}$ by minimizing the total free energy $F_\textrm{total}(z)$ from eq\ \ref{Eq:Total free energy at interface} with respect to $z$.

The input parameters used for obtaining the results presented in Section~\ref{subsec:Wetting behavior} are given in Table~\ref{tbl: Tabulation of input parameters for the model}. The values of $\beta$, $\phi_{0}$, and $N$ are the ones that best fit our experimental data as will be shown in Section~\ref{subsubsec:experimental results of the wetting behavior}. It is important to note that while the best fit values obtained for $\phi_0$ and $N$ happen to agree well with values reported for these quantities in the literature,\cite{2017Lopez,2002Barbero} the independent parameter $\beta$ is determined by the fit. We have not attempted any microscopic theory relating $\beta$, as well as $\phi_0$ and remaining model parameters to the chemical composition of the system. $\gamma$ corresponds to the experimentally measured interfacial tension of a clean water-hexadecane interface and the value of $\nu_a$ is chosen in agreement with literature \cite{2017Lang}. A value of $\delta = 3$ nm as the thickness of the interfacial region has been chosen in accordance to our previous work\cite{2011Isa2,2014Zell}, where it was shown that no dependence of the stretching was observed for larger thicknesses, i.e., the deformation of the shell saturated at  $\delta \geq 3$ nm. We restricted our calculations to CSNPs with shell thickness ranging from 50 nm to 500 nm, as measured in water. While the model imposes no strong limitation on the upper-limit of the shell thickness, one has to be cautious on the choice of the lower limit. If the shell thickness is too small, interactions between the core and the water and oil phases will become predominant over the corresponding interactions with the shell and it must be included in eq~\ref{Eq:Total free energy at interface}. Moreover, the description of the shell as a homogeneous polymer gel also breaks down for thicknesses comparable to the characteristic distance between crosslinks. For a shell thickness greater than 50 nm, all the assumptions in the model can be considered very reasonable and the interactions between core and fluids can be neglected in total free energy (eq~\ref{Eq:Total free energy at interface}).
\begin{table}[htp]
	\caption{Parameters used to model the studied systems. These parameters characterize the solvents, the CSNP chemistry, the thermodynamic state, and the size of polymeric unit. They are unaffected by the radius $R_c$ of the CSNP core and its hydrodynamic radius $R_h$.}
	\label{tbl: Tabulation of input parameters for the model}
    \begin{tabular}{c@{\quad}c@{\quad}c@{\quad}c@{\qquad}c@{\qquad}c@{\qquad}c@{\qquad}c@{\qquad}c@{\qquad}c}
		\hline\hline
Parameter & $\phi_{0}$ & $\chi_w$  & $\chi_o$ & $\beta$ & $T$ & $\nu_a$ & $\delta$ & $\gamma$ & $N$ \\
    &            &  (water)    &   (oil)    &         & [K] & [nm$^3$] & [nm] & [mN/m]  &  \\
          \hline
Value    & 0.0284 & 0.5 & 0.6 & 1.79 & 296 & 0.0328 & 3 & 53.12 & 100\\
		\hline\hline
	\end{tabular}
\end{table}

\section{Results and Discussion}
\label{sec:results and discussion}
\subsection{Size at the Interface and Wetting Behavior}
\label{subsec:Wetting behavior}
\subsubsection{Model predictions}
\label{subsec:model predictions of wetting behavior}
\begin{figure}[htp]
	\centering
	\includegraphics [ scale = 1 ] {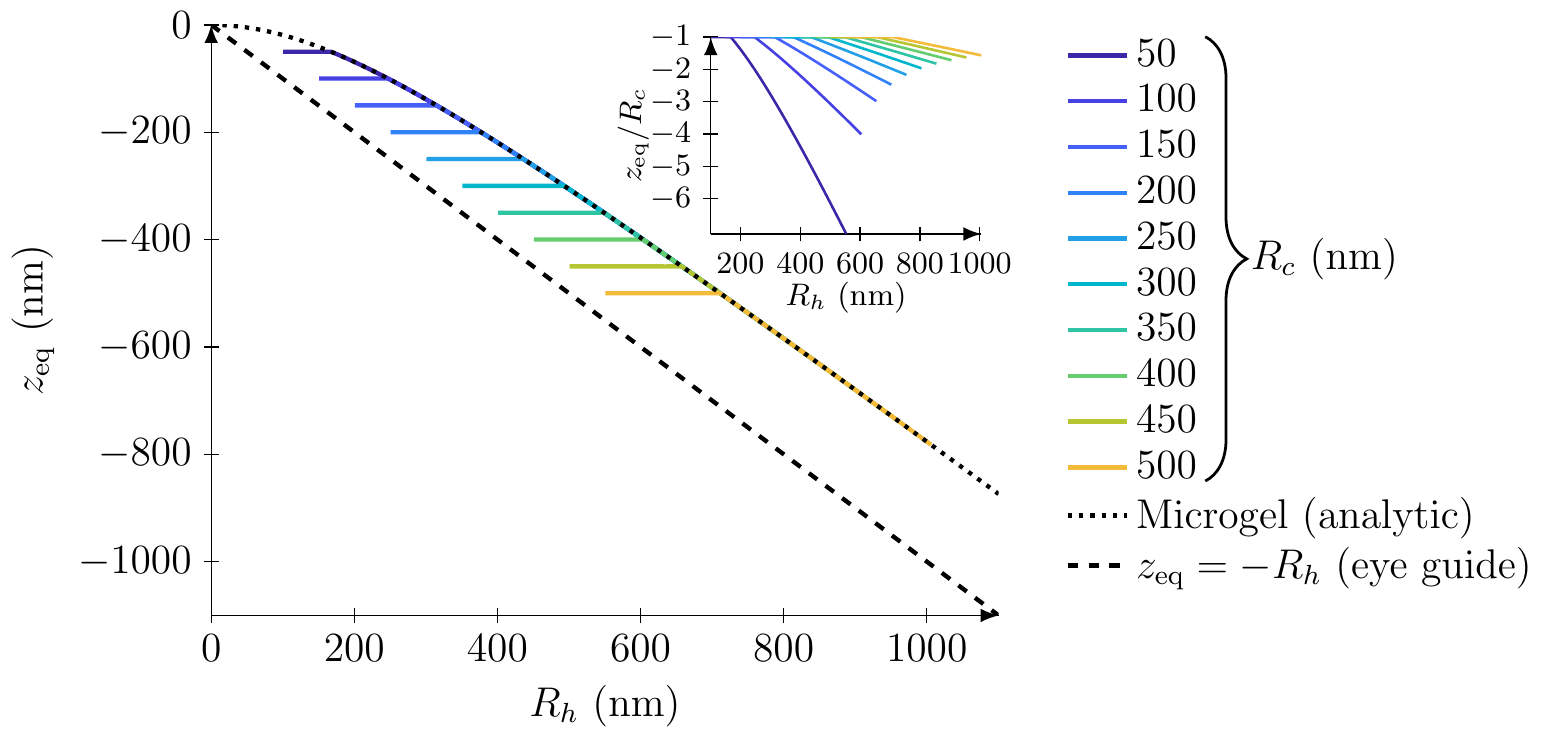}
	\caption{Equilibrium distance of the core $z_\textrm{eq}$ from the interface obtained from the free-energy-minimization model as a function of the hydrodynamic radius of the CSNP in water $R_h$. The dotted line corresponds to the solution of eq~\ref{Eq:Solution of z equilibrium}, which describes the equilibrium position of a microgel, i.e. in the absence of a silica core. Inset: $z_\textrm{eq}$ scaled by $R_c$ as a function of $R_h$.}
	\label{fig:equilibrium height and radius of the CSNP for different core sizes}
\end{figure}

We start by presenting the results of the model described above for particles of different core sizes as a function of shell thickness to provide a framework to rationalize the experimental data shown in the following section. Figure~\ref{fig:equilibrium height and radius of the CSNP for different core sizes} shows the equilibrium height $z_\textrm{eq}$ of the CSNP obtained by minimizing the free energy in eq~\ref{Eq:Total free energy at interface} for different cores sizes and shell thickness ranging from 50 nm to 500 nm, corresponding to particles with hydrodynamic radii in water $R_{c}+50$ nm $\leq R_{h}\leq R_{c}+500$ nm.
 The values of $z_\textrm{eq}$ are negative because the center of the core is located in the water phase. As expected, all the CSNPs, irrespective of their core and shell size, prefer to be in contact with the interface to remove some of the water-oil interfacial area. However, two notable features can be observed. Until a critical value of $R_{h}$, which is weakly dependent on $R_{c}$, the core is always found in a state wherein it just touches the interface, i.e., $z_\textrm{eq} = - R_{c}$. An increase in $R_{h}$ beyond a critical value causes instead $z_\textrm{eq}$ to decrease continuously and in the same manner for all CSNPs, irrespective of their core radius.

\begin{figure*}[htp]
	\centering
	\includegraphics [ scale = 1 ] {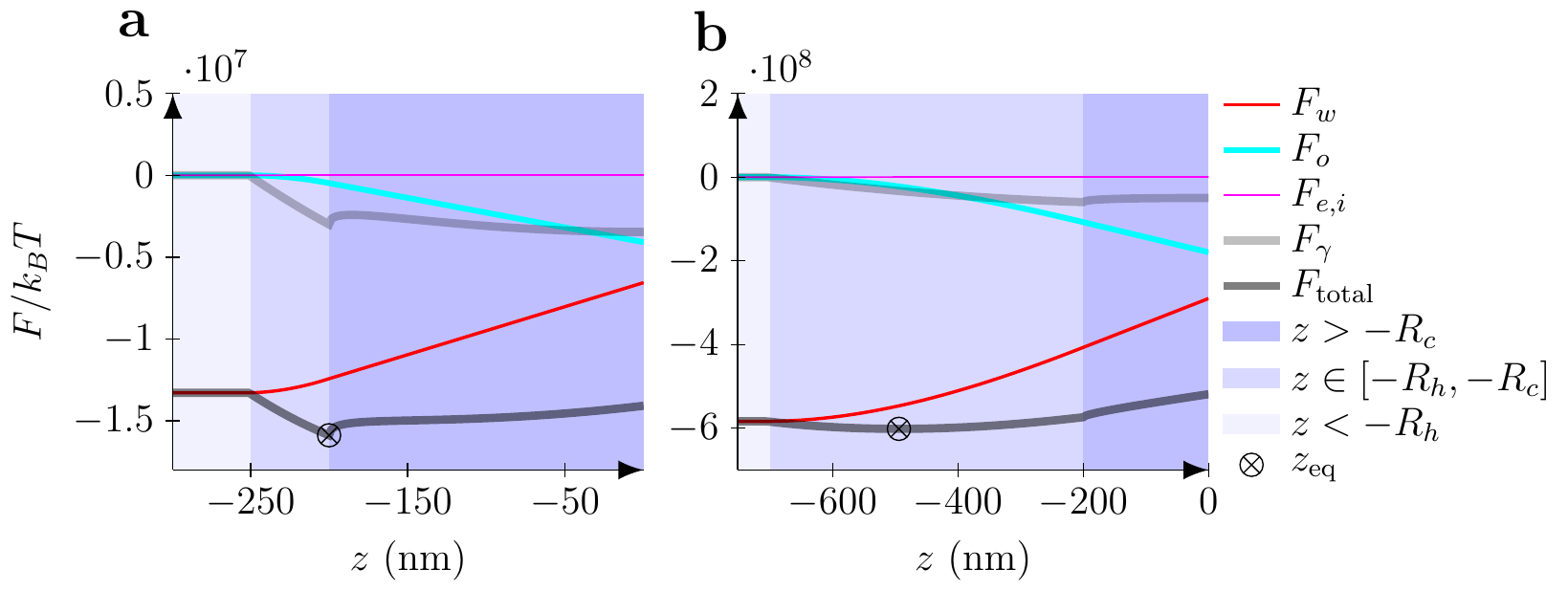}	
	\caption{Free-energy components and total free energy as a function of the position $z$ of CSNPs at the interface. The data in \textbf{(a)} and \textbf{(b)} correspond to a CSNP with a shell thickness of 50 nm and 500 nm respectively, while the core radius is $R_c=200$ nm for both of them.}
	\label{fig:free energy component variation with z}
\end{figure*}

 The inset of Fig.~\ref{fig:equilibrium height and radius of the CSNP for different core sizes} shows more explicitly the $z_\textrm{eq} = - R_{c}$ behavior of CSNPs until the critical value of $R_{h}$, where $z_\textrm{eq}$ is scaled by $R_{c}$. This is the energetically favored state of the particle for these values of shell thickness because the minimum of $F_\textrm{total}$ occurs at the same position as the minimum of $F_{\gamma}$ (e.g. see Fig.~\ref{fig:free energy component variation with z}). Correspondingly, the minimum of the latter quantity always occurs at a $z$-position at which a given CSNP removes the maximum area of the interface. As seen from Fig.~\ref{fig:free energy component variation with z}, the elastic energy penalty coming from the shell stretching, $F_{e,i}$, has a negligible contribution when compared to the other free-energy components. Hence, $z_\textrm{eq} = -R_{c}$ until the energy gained from the interfacial area removal compensates for the energy penalty coming from the exposure of the shell to the oil phase, i.e.,
\begin{equation*}
z_\textrm{eq} = -R_c \qquad\forall\qquad
  \left|F_{\gamma}(-R_{c})\right|> \left|F_{w}(-R_{h})\right|-\left|F_{w}(-R_{c})+F_{o}(-R_{c})\right|.
\end{equation*} Additionally, the expression above emphasizes that the range of shell thicknesses for which $z_\textrm{eq} =-R_{c}$ can be tuned by optimizing the solvent quality of the two liquid phases. However, $F_w$ becomes sufficiently large in the second regime where all CSNPs behave identically, resulting in a shallow minimum in the total free energy for $z<-R_c$. This is equivalent to state that the interface no longer feels the presence of the core, and hence the CSNP behaves as a pure microgel, i.e., a pure PNIPAM cross-linked particle without the silica core. We henceforth derive an analytic expression for $z_\textrm{eq}$ as a function of particle properties in the case of a pure microgel. This is done by solving
\begin{equation}\label{Eq:microgel equations}
\frac{dF_\textrm{total}(z)}{dz} = 0,
\end{equation}
where in the expression for $F_\textrm{total}(z)$ given by eq~\ref{Eq:Total free energy at interface} the core radius $R_c$ is set to zero in eqs~\ref{Eq:radius of particle at interface}, \ref{Eq:volume of shell at interface before stretching}, \ref{Eq:volume of shell at interface after stretching}, \ref{Eq:volume of shell in water phase} and \ref{Eq:volume of shell in oil phase} to account for the absence of the core. This results in the following simplified relationships:
\begin{align}
R_{i} &= \beta \sqrt{R_{h}^2-z^2}\\
V_{i,b} &= \pi \delta (R_{h}^2-z^2)\\
V_{i} &= \pi \delta \beta^2 (R_{h}^2-z^2)\\
V_{w} &= \frac{\pi}{3}(R_h-z)^2(2R_h+z)-V_{i,b} \\
V_{o} &= \frac{\pi}{3} \left(\frac{\phi_{w}}{\phi_{o}}\right)(R_{h}+z)^2 (2R_{h}-z)
\end{align}
By taking the derivatives of these expressions with respect to $z$ we obtain the following:
\begin{align}
\frac{dR_{i}}{dz} &= \frac{-\beta z}{\sqrt{R_{h}^2-z^2}} \label{Eq:differential1}\\
\frac{dV_{i,b}}{dz} &= -2\pi \delta z \\
\frac{dV_{i}}{dz} &= -2\pi \delta z \beta^2\\
\frac{dV_{w}}{dz} &= -\pi (R_h^2-z^2) + 2\pi \delta z\\
\frac{dV_{o}}{dz} &= \pi (R_h^2-z^2) \left(\frac{\phi_{w}}{\phi_{o}}\right)
\label{Eq:differential2}
\end{align}
Substitution of eqs~(\ref{Eq:differential1})--(\ref{Eq:differential2}) into eq~\ref{Eq:microgel equations} and further simplifications yield
\begin{equation}\label{Eq:quadratic expression for z height}
\left[f_{w}-f_{o}\frac{\phi_{w}}{\phi_{o}}\right]z^2 - 2(f_{e,i}\delta\beta^2-f_{w}\delta-\gamma\beta^2\nu_{a})z - \left[f_{w}-f_{o}\frac{\phi_{w}}{\phi_{o}}\right]R_{h}^2 = 0.
\end{equation}
Since this is a quadratic expression in $z$, it has two solutions of which the only relevant one for our case is given by
\begin{equation}\label{Eq:Solution of z equilibrium}
z_\textrm{eq} = \frac{-B + \sqrt{B^2-4AC}}{2A},
\end{equation}
where
\begin{align*}
A &= f_{w}-f_{o}\frac{\phi_{w}}{\phi_{o}}, \\
B &= -2(f_{e,i}\delta\beta^2-f_{w}\delta-\gamma\beta^2\nu_{a}),\textrm{ and} \\
C &= -A R_{h}^2.
\end{align*}
This expression is shown by the dotted black line in Fig.~\ref{fig:equilibrium height and radius of the CSNP for different core sizes} together with results of the model data for the input parameters reported in Table~\ref{tbl: Tabulation of input parameters for the model}. As expected, the analytic expression captures the regime above the critical shell thickness for which the core leaves the interface. As we will show later, this expression can be used as a guide to synthesize CSNPs with a well-defined size at the interface, if some key physical parameters of the shell of the CSNP are known.

\begin{figure}[htp]
	\centering
	\includegraphics [ scale = 1 ] {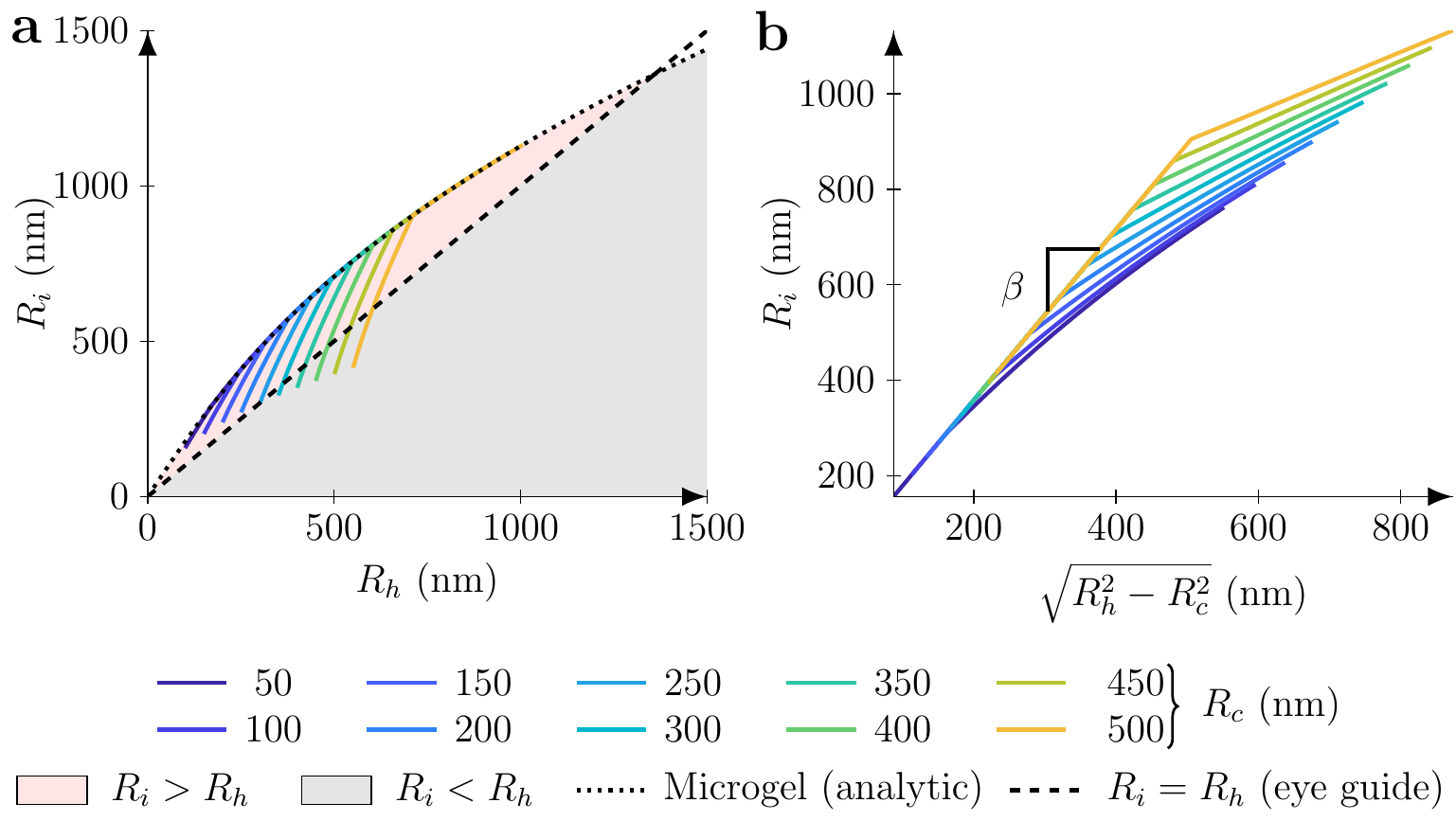}	
	\caption{\textbf{(a)} Equilibrium radius of CSNPs at interface, $R_{i}$, as a function of the radius of the CSNPs in the water phase $R_{h}$. The dotted line corresponds to the analytical prediction of $R_{i}$ for a microgel. The pink- and the gray-shaded areas represent regions for which $R_{i} \geq R_{h}$ and $R_{i} \leq R_{h}$, respectively. \textbf{(b)} $R_{i}$ as a function of $\sqrt{R_{h}^2-R_{c}^2}$. The slope of the linear regime is the stretching parameter $\beta$ of the CSNP used in the model.}
	\label{Fig: Model eq. radius variation}
\end{figure}

In addition to the vertical position relative to the interface, the model also provides us with the value of the equilibrium interface radius, $R_i$. Figure~\ref{Fig: Model eq. radius variation}a shows $R_i$ as a function of $R_h$ for the same range of input parameters.

The observed behavior of $R_{i}$ clearly demonstrates the direct effect of the $z_\textrm{eq}$ on the CSNP's interfacial radius. In addition, it also shows that beyond a critical value of $R_{h}$, which corresponds to the crossover between the dotted and dashed black lines in Fig.~\ref{Fig: Model eq. radius variation}a, all CSNPs and microgels have $R_{i}<R_{h}$, in agreement with the experimentally reported values for micron-sized microgels.\cite{2016Kwok}.

For the regime where the core just touches the interface, $R_i$ is equal to $\beta \sqrt{R_{h}^2-R_{c}^2}$. Hence, by plotting $R_i$ as a function of $\sqrt{R_{h}^2-R_{c}^2}$ we should observe a linear-regime, as demonstrated in Fig.~\ref{Fig: Model eq. radius variation}b. In addition, this plot also allows us to observe two key features. First, it tells us that by fitting the region of $R_{i}$ which varies linearly with $\sqrt{R_{h}^2-R_{c}^2}$ we can extract the stretching parameter of the shell $\beta$, which otherwise is difficult to extract experimentally without knowledge of the exact position of core with respect to the interface. Moreover, it also points out that, beyond the linear regime, $R_{i}$ increases rather slowly with increasing $R_{h}$ for a fixed value of $R_{c}$. In other words, one could attain the same value of $R_{i}$ by using a CSNP with a large core and small shell thickness instead of using a small core with very large shell thickness. These results could be particularly relevant for colloidal lithography applications, e.g., for the fabrication of silicon nanowires with a desired diameter and spacing using soft-templates \cite{2016Rey}. For our present purpose, this provides us with an optimal method to estimate both the position of the core and the stretching parameter $\beta$ from our experimental data.

\subsubsection{Experiments}
\label{subsubsec:experimental results of the wetting behavior}
The results of the model are tested by experimentally measuring  $R_{i}$ for a systematic series of CSNPs. We investigated two sets of particles with core radii, $R_{c}$, of 63 nm (C1) and 176 nm (C2-C3). For each value of $R_{c}$,  four different values of shell thickness $S_{t}$ were investigated (see Table~\ref{tbl: Tabulation of the particle sizes used in the article}), where $S_{t}=R_{h} - R_{c}$ is defined as the difference between the hydrodynamic radius at 23 \textdegree{}C and the core radius.
\begin{table}[htp]
	\small
	\caption{Characterization of the studied CSNPs with two different core radii $R_c$. Hydrodynamic radius $R_h$ at 23 \textdegree{}C in water; $S_t=R_h-R_c$ is the corresponding shell thickness.}
	\label{tbl: Tabulation of the particle sizes used in the article}
	\begin{tabular*}{0.48\textwidth}{@{\extracolsep{\fill}}llll}
		\hline\hline
		$R_c$ [nm] & CSNP & $R_h$ [nm] & $S_t$ [nm] \\
		\hline
		\multirow{4}{*}{63} & C1S1 & $117\pm9$ & 54 $\pm 9$ \\
		& C1S2 & $135\pm9$ & $72\pm9$ \\
		& C1S3 & $183\pm14$ & $120\pm14$ \\
		& C1S4 & $208\pm13$ & $145\pm13$ \\ \hline
		\multirow{4}{*}{176} & C2S1 & $297\pm19$ &  $121\pm19$ \\
		& C3S1 & $325\pm31$ & $149\pm31$\\
		& C2S2 & $351\pm52$ & $175\pm52$ \\
		& C3S2 & $510\pm114$ & $334\pm114$ \\
		\hline\hline
	\end{tabular*}
\end{table}

\begin{figure}[htp]
	\centering
	\includegraphics [scale = 1] {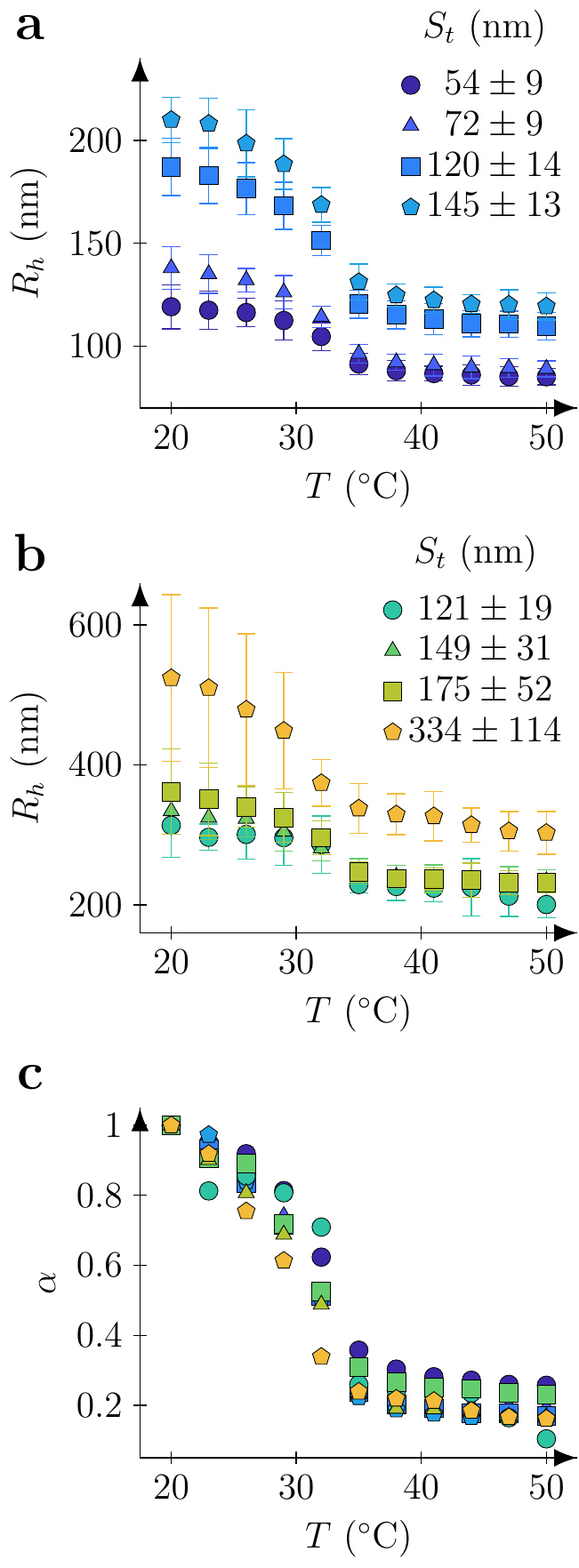}
	\caption{Hydrodynamic radius $R_h$ as a function of temperature $T$ for \textbf{(a)} $R_c=$ 63 nm and \textbf{(b)} 176 nm cores as obtained from DLS measurements. The shell thickness is defined as $S_t=R_h-R_c$ at 23 \textdegree{}C. \textbf{(c)} Deswelling parameter $\alpha$ as a function of temperature for both cores, the symbols are the same as used for \textbf{(a)} and \textbf{(b)}.}
	\label{fig: variation of the hydrodynamic radius and deswelling ratio as a function of temperature}
\end{figure}

Before presenting the results of the $R_{i}$ measurements, we need to demonstrate that the assumption made in the model that all CSNPs have effectively the same cross-linking density independent of their shell thickness is justified. In fact, as highlighted in the schematic representation of the CSNP in Fig.~\ref{fig: schematic representation of csnp in bulk and water-oil interface}a, the PNIPAM shell encapsulating the core of the CSNP has a radial cross-linking density profile. This is a direct consequence of the synthesis procedure used \cite{2004Stieger}. A comparison of the cross-linking densities between different particle batches can be made by measuring the temperature-dependent behavior of their hydrodynamic radius $R_h$. The variation of $R_h$ as a function of temperature $T$ is shown in Figs.~\ref{fig: variation of the hydrodynamic radius and deswelling ratio as a function of temperature}a and \ref{fig: variation of the hydrodynamic radius and deswelling ratio as a function of temperature}b. 
For the sake of clarity, the CSNPs with different $R_c$ values are plotted as individual subfigures. With increasing temperature, water is expelled from the PNIPAM shell resulting in a decrease of the hydrodynamic radius, which eventually reaches a plateau at around 35 \textdegree{}C when PNIPAM-PNIPAM interactions are preferred over the PNIPAM-water interactions\cite{2016Rauh}. This temperature is slightly above the one established for bulk PNIPAM\cite{Halperin2015}. Figure~\ref{fig: variation of the hydrodynamic radius and deswelling ratio as a function of temperature}c shows the variation of the deswelling parameter, $\alpha$, defined as the ratio of the volume of the polymer shell of a CSNP at a given temperature $T$ by the volume of the polymer shell at 20 \textdegree{}C, i.e., $\alpha = (R_{h,T}^3-R_{c}^3)/(R_{h,20^{\circ}\textrm{C}}^3-R_{c}^3)$, in overall agreement with theoretical expectations\cite{Halperin2011}. The deswelling parameter reaches a plateau value, which is nearly the same for all particles, hence supporting the assumption that all particles have the same effective cross-linking density.

\begin{figure}[htp]
	\centering
	\includegraphics[width = 65mm, height=65mm]{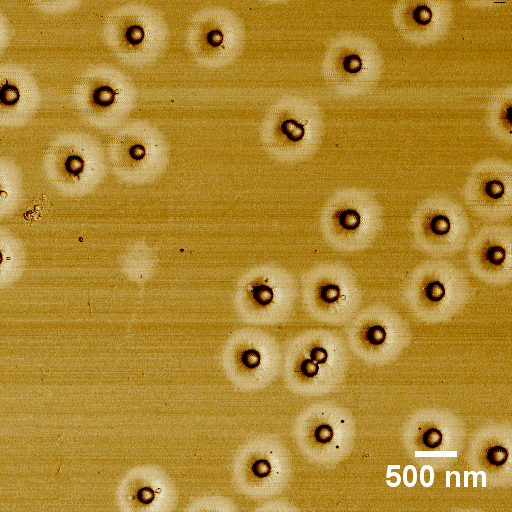}
	\caption{AFM phase image of the C1S4 CSNPs after deposition on a silicon wafer from the water-hexane interface. The C1S4 CSNPs have a core radius of 63 nm and shell thickness of $145\ \textrm{nm}\pm13$ nm in water.}
	\label{fig:phase image of ar108}
\end{figure}

We used two different approaches to characterize the size of the CSNP at the interface, the details of which are discussed in Sections~\ref{subsubsec: depositions and afm imaging} and \ref{subsubsec: dynamics and sizing at the interface}. In the first approach, particle sizes were measured from AFM phase images of the CSNPs deposited from the water-hexane interface onto a silicon wafer \cite{2017Scheidegger}. An example of a phase image for CSNPs with a core radius of 63 nm and shell thickness of $145\ \textrm{nm}\pm13$ nm in water phase is shown in Fig.~\ref{fig:phase image of ar108}. In the phase image, the part of the shell which is stretched-out at the interface is seen after deposition as a bright region surrounding the darker inner ring, which approximately marks the outer edge of the silica core. By measuring the size of 55 -- 146 particles for each batch, we can determine their average size at the interface. While this procedure allowed us to reliably characterize the size of the CSNPs with core radius of 63 nm, it could not be used for CSNPs with core radius of 176 nm. In order to understand this limitation, we need to review the mechanism of the deposition process at a single-particle scale. Consider for instance the deposition of a particle such as the one depicted in Fig.~\ref{fig: schematic representation of csnp in bulk and water-oil interface}b. A silicon wafer which is initially immersed in the water phase, is slowly pulled upwards to meet the particle at the interface. When the silicon wafer eventually comes in contact with the periphery of the shell at the interface, the latter attaches to the substrate. Since the outer edge of the shell has a fixed distance from the periphery of the core in the plane of the interface, it has to take-up a shape which allows for both the core and part of the shell, which is exposed to the water phase, to deposit onto the substrate. This would result in a net decrease in the shell thickness measured after deposition. The extent of error in the measured size of the CSNP increases with an increase in the core radius and the height $z$.

\begin{figure}[htp]
	\centering
	\includegraphics {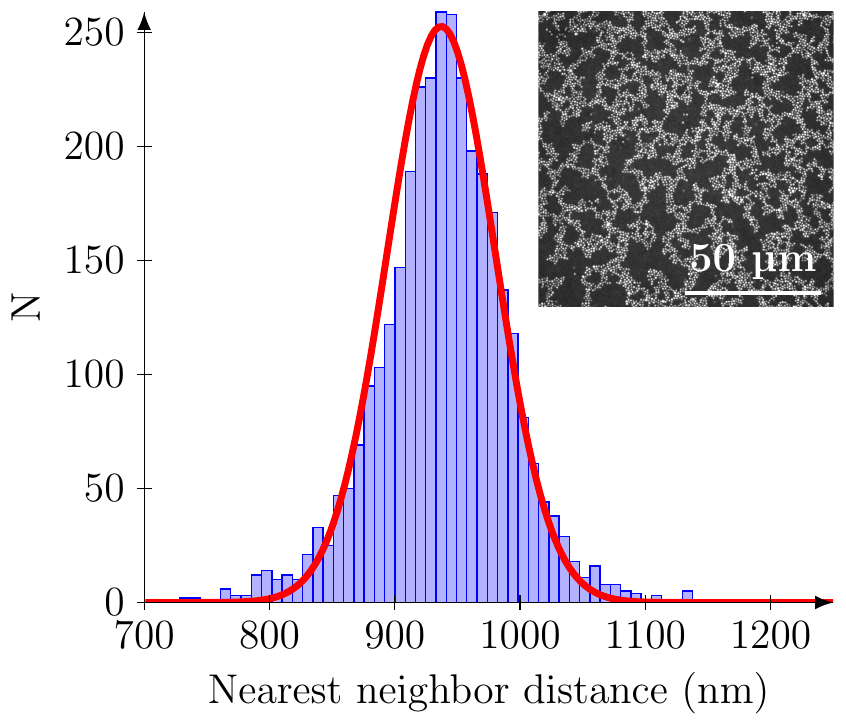}
	\caption{Histogram of the nearest-neighbor distances between C3S1 CSNPs in a Gibbs monolayer formed at the water-hexadecane interface fitted with a normal distribution (solid red line). Inset: confocal image of the monolayer. The data and image corresponds to C3S1 CSNP with a core radius of 176 nm and shell thickness of $149\ \textrm{nm}\pm31$ nm in water.}
	\label{fig:Image and histogram of Ar174}
\end{figure}

For this reason,  $R_{i}$ of the C2/C3 CSNPs with core radius of 176 nm was estimated by determining the nearest-neighbor distances from Gibbs monolayers of the particles at the water-hexadecane interface. A confocal microscopy image of a monolayer formed by C3S1 CSNPs with shell thickness (in water) of $149\ \textrm{nm}\pm31$ nm is shown in the inset to Fig.~\ref{fig:Image and histogram of Ar174}.
The particles form aggregated networks due to attractive capillary forces between them, wherein the particles shells are in contact\cite{2016Huang,2018Vasudevan}. The histogram of the nearest-neighbor distances obtained from more than 1100 particles is fitted with a normal distribution to obtain the average particle radius at the interface, $R_i$, and its standard deviation (see Fig.~\ref{fig:Image and histogram of Ar174}). Due to their compressibility, when the shells come in contact, the measured values of $R_i$ may slightly underestimate the unperturbed values.

\begin{figure}[htp]
	\centering
	\includegraphics {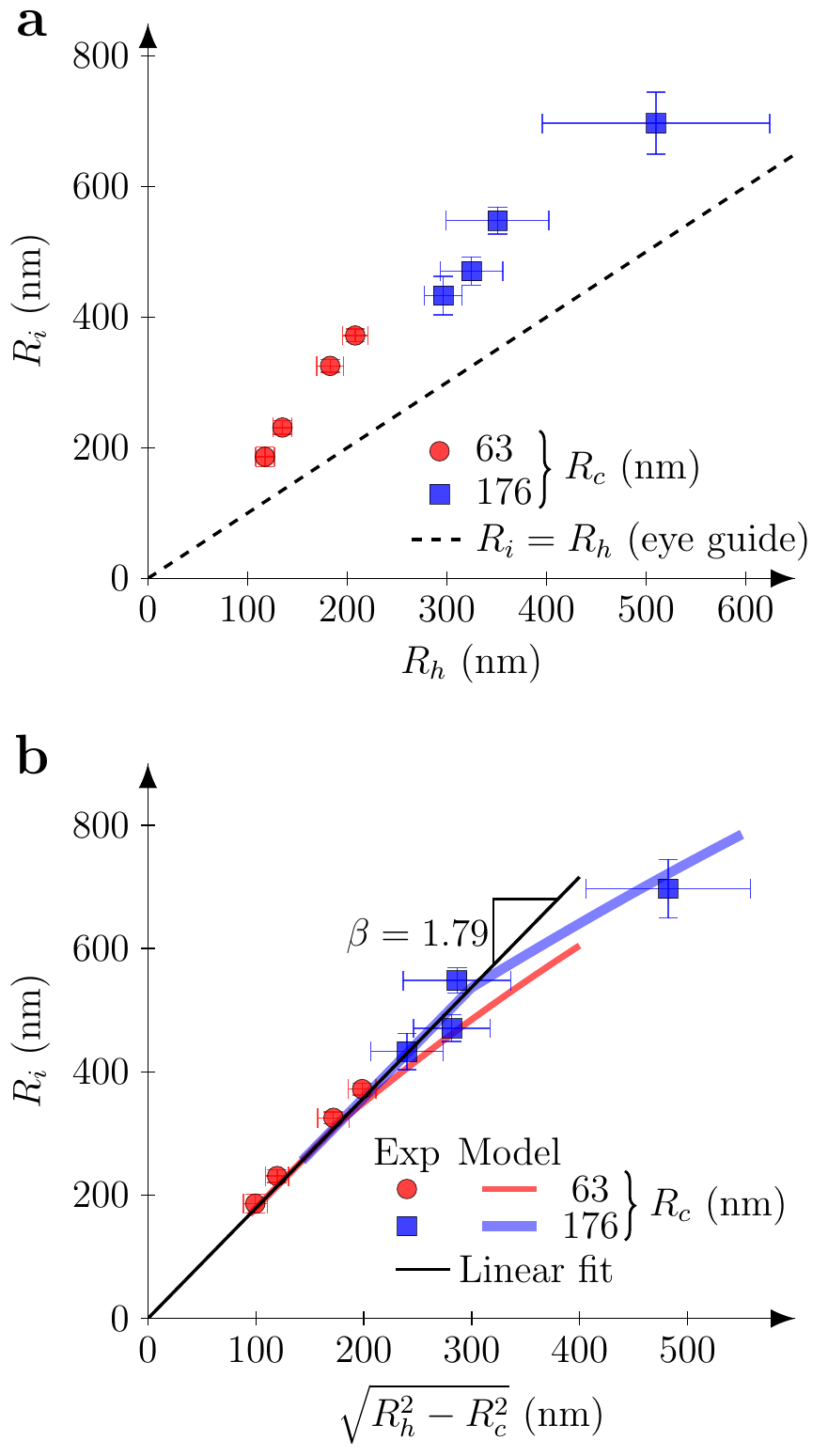}
	\caption{\textbf{(a)} CSNP radius at interface $R_{i}$ versus the particle radius in bulk water $R_{h}$ at 23 \textdegree{}C. \textbf{(b)} Experimental data of the CSNP radius at interface $R_{i}$ as a function of $\sqrt{R_{h}^2-R_{c}^2}$. The solid black line is a fit to the data in the linear regime, whose slope provides us with parameter $\beta$, while the red and blue lines are our model predictions for $R_{c} = $ 63 nm and 176 nm, respectively.}
	\label{fig:RadIntvsRadBulkAndRadTilde}
\end{figure}

Figure~\ref{fig:RadIntvsRadBulkAndRadTilde}a shows the measured values of $R_i$ as a function of the particle radius in bulk water at 23 \textdegree{}C. The particle radius at the interface is much larger than the bulk radius for all CSNPs.
Following the indications of the model, we plot the experimental behavior of $R_{i}$ as a function of $\sqrt{R_{h}^2-R_{c}^2}$ in  Fig.~\ref{fig:RadIntvsRadBulkAndRadTilde}b. It is clear from the data that for all particles except one (the last blue data point in the plot), $R_{i}$ changes linearly with $\sqrt{R_{h}^2-R_{c}^2}$. By comparing the data with the model predictions, the measurements indicate that all these particles adopt a configuration wherein the core just touches the interface. The last blue data point corresponds to a case where the core is below the interface. By fitting the linear part of the experimental data, we extract the stretching parameter $\beta$ for the shell of our CSNPs to be 1.79. This value of the stretching parameter is in agreement with numerical studies\cite{2016Mehrabin}.
\subsection{Dynamics}
\label{subsec:dynamics}
\subsubsection{Bulk water}
\label{subsubsec:bulk dynamics}
\begin{figure}[t]
	\centering
	\includegraphics {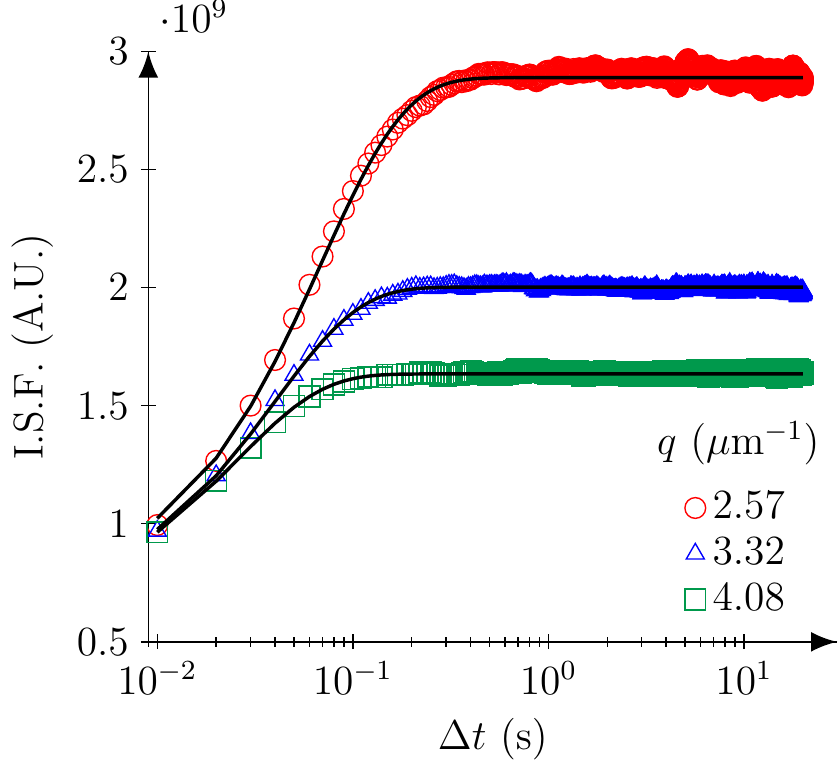}
	\caption{Intermediate scattering function (I.S.F.) versus lag-time ${\Delta t}$ for three different wavenumbers $q$. The data shown is for the C1S2 CSNP with core radius $R_c=$ 63 nm and a shell thickness of $S_t=72\ \textrm{nm}\pm9$ nm. The empty symbols are the experimental data and the black solid lines are fits to eq~\ref{eq: isf fitting function}.}
	\label{fig: isf behavior for ar116 csnp}
\end{figure}

\begin{figure}[htp]
	\centering
	\includegraphics {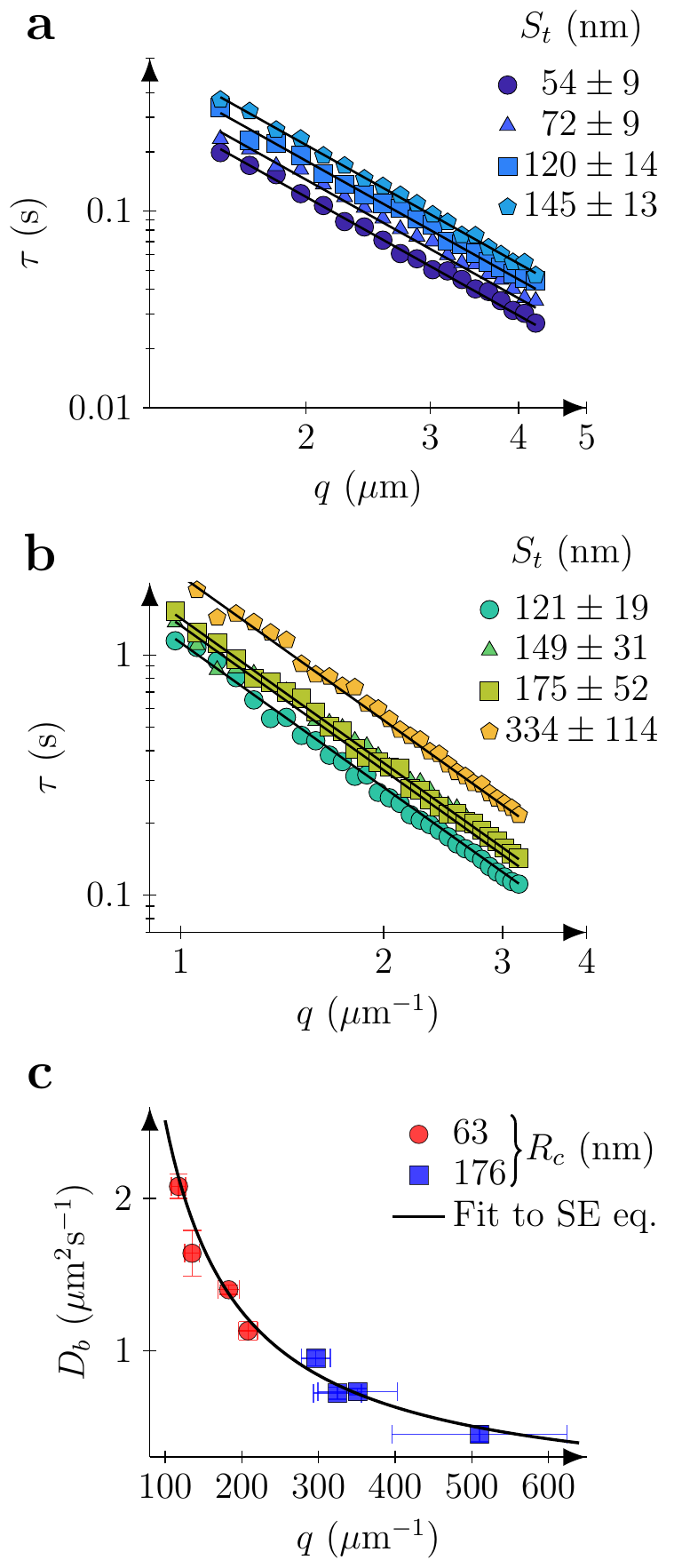}
	\caption{\textbf{(a)} and \textbf{(b)} show the $\tau$ vs $q$ behavior for CSNPs with core radius of 63 nm and 176 nm, respectively. The solid black lines are fite of the experimental data to eq~\ref{Eq:tau dependence on q}. \textbf{(c)} Bulk diffusion coefficient dependence on hydrodynamic radius $R_{h}$ of the CSNPs, the solid black line is a fit of the experimental data (filled symbols) to the SE equation.}
	\label{fig: Tau vs q behavior and bulk diffusion coefficient.}
\end{figure}

As discussed in the Introduction, the dynamics of the particles at the interface is coupled to their shape. Before measuring the interfacial dynamics, we first present data on the particle dynamics in bulk water as a function of the core size and shell thickness, as obtained by DDM. Figure~\ref{fig: isf behavior for ar116 csnp} shows the typical behavior of the intermediate scattering function (I.S.F.) obtained from a DDM measurement. The experimental data is fitted to the function in eq~\ref{eq: isf fitting function} to obtain the characteristic relaxation time $\tau$ for each value of the wavenumber $q$ \cite{2008Cerbino}, \begin{equation}\label{eq: isf fitting function}
\mathrm{I.S.F}(q,\Delta t) = A(q)\left[1-\exp\left(-\Delta t/ \tau\right)\right]+B(q),
\end{equation} where $A(q)$ and $B(q)$ are treated as $q$-dependent fitting parameters.
The $\tau$ values thus obtained for CSNPs with core radius of 63 nm and 176 nm are shown respectively in Figs.~\ref{fig: Tau vs q behavior and bulk diffusion coefficient.}a and \ref{fig: Tau vs q behavior and bulk diffusion coefficient.}b. For particles exhibiting normal Brownian motion, $\tau$ has the following dependence on the wavenumber $q$:\begin{equation}\label{Eq:tau dependence on q}
\tau = \frac{1}{D_{b}q^2}.
\end{equation} By fitting the experimental data to the above expression (solid black lines in Fig.~\ref{fig: Tau vs q behavior and bulk diffusion coefficient.}a and Fig.~\ref{fig: Tau vs q behavior and bulk diffusion coefficient.}b), we obtain the diffusion coefficient $D_{b}$ of the CSNPs in bulk water phase, shown in Fig.~\ref{fig: Tau vs q behavior and bulk diffusion coefficient.}c. As expected, the diffusivity of these particles at dilute concentrations in water is well captured by the Stokes-Einstein (SE) equation \cite{2003Routh} (solid black line in Fig.~\ref{fig: Tau vs q behavior and bulk diffusion coefficient.}c), $D_{b} = k_{\textrm{B}}T/\left(6\pi\eta R_{h}\right)$, wherein $k_{B}$ and $T$ are, respectively, the Boltzmann constant and the absolute temperature and the bulk hydrodynamic radius, $R_{h}$, of the CSNP is used as the characteristic size of the particle. The viscosity obtained from the fit, $\eta = 8.8\times10^{-4}$ Pa$\cdot$s, corresponds to that of water at 23 \textdegree{}C. This finding confirms that the dynamics of these particles at dilute conditions in water is identical to that of hard spheres.

\subsubsection{Water-oil interface}
\label{subsubsec:dynamics interface}
\begin{figure}[htp]
	\centering
	\includegraphics[scale = 1]{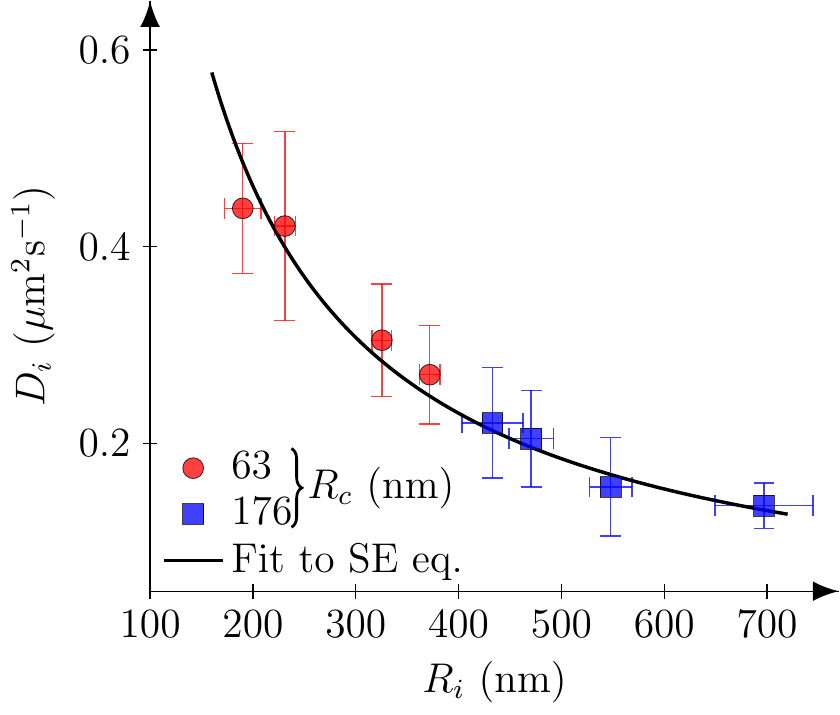}
	\caption{Diffusion coefficient $D_{i}$ of CSNPs at the water-hexadecane interface as a function of the radius of the particle at the interface $R_{i}$. The solid black line is a fit of the data to the SE equation.}
	\label{Fig: diffusion coefficient of csnps at interface}
\end{figure}

After identifying the position of the CSNPs relative to the interface and their interfacial cross-sectional radius, we proceed here to measure their diffusivity at the interface and compare it to the bulk values. Figure~\ref{Fig: diffusion coefficient of csnps at interface} shows the diffusion coefficient $D_{i}$ of the CSNP at the interface as a function of their radius at the interface $R_{i}$. 
In analogy with the bulk diffusivity measurements, both the particle diffusivity and size at the interface were determined independently. The details of the method used for extracting the particle diffusivity were discussed in Section~\ref{subsubsec: dynamics and sizing at the interface}. From the data, it is evident that particles diffuse much more slowly at the interface when compared to bulk measurements (Fig.~\ref{fig: Tau vs q behavior and bulk diffusion coefficient.}c). We find that the diffusivity behavior can be captured very well by using the SE equation (black solid line in Fig.~\ref{Fig: diffusion coefficient of csnps at interface}) provided that the particle radius at the interface is used as the characteristic radius of the particle. From the SE relation, we extract an effective viscosity experienced by the particle as a fitting parameter, obtaining a value of $2.3\times10^{-3}$ Pa$\cdot$s, in between the water and hexadecane bulk viscosities.

These findings are very interesting. In particular, it is rather surprising that the diffusivity behavior of all the particles is well captured by the SE equation with a single value of the fitting parameter (i.e. the same value of effective viscosity). In fact, from the results discussed in the previous section, we know that for each kind of CSNP the ratios of the particle surfaces exposed to water and oil phases are different because $z_\textrm{eq} = -R_{c}$. Hence, one would expect the effective viscosity experienced by each particle batch to be different and a function of the specific particle shape. This experimental finding emphasizes a marked difference with the behavior of hard spheres, for which the viscous drag at the interface is a function of the surface ratio exposed to the two fluids, i.e. their contact angle \cite{2016Dorr,1995Danov,2006Fischer}. Albeit the complexity of the problem requires further investigations, the data in Fig.~\ref{Fig: diffusion coefficient of csnps at interface} suggests that, in first approximation, the dominant factor in the drag experienced by the CSNPs  does not come from the bulk phases but rather from the interface.

\section{Conclusions}
\label{sec:Conclusions}
To summarize, we have systematically studied the behavior of CSNPs with different silica core and PNIPAM shell sizes to characterize their wetting, size and diffusivity at a water-oil interface. We both predict from a model and experimentally observe that the interfacial size of the CSNP is strongly dependent on both the core size and the shell thickness. This dependence comes about from the equilibrium position of the core with respect to the interface. In particular, our model is consistent with the data and finds that for a given core size and shell thickness below a critical value, the core sits in a state wherein it just touches the interface while being immersed in the water phase. However, increasing shell thickness beyond the critical value causes the core to detach from the interface and sit deeper into the water phase with no memory of its presence at the interface. Our results further demonstrate that the position of the core relative to the interface plane can be obtained by simply measuring the particle dimensions at the interface (provided that the shell properties are known). The latter quantity is experimentally much more accessible than the former one, which remains elusive to measure for small colloids and escapes many measurement techniques, including freeze-fracture shadow-casting (FreSCa) for highly hydrophilic CSNPs\cite{2011Isa1}. This finding will be useful for applications where controlled assemblies of CSNPs at liquid-liquid interfaces are of interest, for instance for surface-enhanced Raman scattering, sensing and nanopatterning\cite{2009Alvarez,2016Yang,2011Karg}. Concerning dynamics, we find that, although these CSNPs are mostly immersed in water, they experience a viscous drag at the interface much larger than the one expected from bulk water viscosity. Although more detailed studies are required to capture the complex hydrodynamics, our data suggest that the dominant factor for the viscous dissipation does not come from bulk phases but rather from the interface. The presence of aggregation at the interface, as for instance displayed in the inset of Fig.~\ref{fig:Image and histogram of Ar174}a, shows that, upon adsorption, the shell induces attractive capillary forces stemming from heterogeneities of the three-phase contact line\cite{2016Huang,2018Vasudevan}. These heterogeneities may be responsible for the increased drag at the interface, as already hypothesized for non-deformable spheres \cite{2015Boniello}. Moreover, unlike their hard-sphere counterparts, the drag experienced by the CSNPs appears to be independent of their position relative to the plane of the interface and to simply scale with the interfacial radius. These findings highlight the subtle interplay between CSNP conformation, position, and dynamics at the interface and emphasize the dominant role played by the shell, prompting further investigations on the potential use of CSNPs as sensitive interfacial tracers.



\begin{thebibliography}{38}
	
	\bibitem[Plamper and Richtering(2017)Plamper, and Richtering]{2017Plamper}
	Plamper,~F.~A.; Richtering,~W. Functional Microgels and Microgel Systems.
	\emph{Acc. Chem. Res.} \textbf{2017}, \emph{50}, 131--140
	\bibitem[Yunker \latin{et~al.}(2014)Yunker, Chen, Gratale, Lohr, Still, and
	Yodh]{2014Yunker}
	Yunker,~P.~J.; Chen,~K.; Gratale,~M.~D.; Lohr,~M.~A.; Still,~T.; Yodh,~A.~G.
	Physics in ordered and disordered colloidal matter composed of
	poly(N-isopropylacrylamide) microgel particles. \emph{Rep. Prog. Phys.}
	\textbf{2014}, \emph{77}, 056601
	\bibitem[Karg(2016)]{2016Karg}
	Karg,~M. Functional Materials Design through Hydrogel Encapsulation of
	Inorganic Nanoparticles: Recent Developments and Challenges. \emph{Macromol.
		Chem. Phys.} \textbf{2016}, \emph{217}, 242--255
	\bibitem[Destribats \latin{et~al.}(2011)Destribats, Lapeyre, Wolfs, Sellier,
	Leal-Calderon, Ravaine, and Schmitt]{2011Destribats}
	Destribats,~M.; Lapeyre,~V.; Wolfs,~M.; Sellier,~E.; Leal-Calderon,~F.;
	Ravaine,~V.; Schmitt,~V. Soft microgels as Pickering emulsion stabilisers:
	role of particle deformability. \emph{Soft Matter} \textbf{2011}, \emph{7},
	7689--7698
	\bibitem[Rauh \latin{et~al.}(2017)Rauh, Rey, Barbera, Zanini, Karg, and
	Isa]{2016Rauh}
	Rauh,~A.; Rey,~M.; Barbera,~L.; Zanini,~M.; Karg,~M.; Isa,~L. Compression of
	hard core-soft shell nanoparticles at liquid-liquid interfaces: influence of
	the shell thickness. \emph{Soft Matter} \textbf{2017}, \emph{13},
	158--169
	\bibitem[Style \latin{et~al.}(2015)Style, Isa, and Dufresne]{2015Style}
	Style,~R.~W.; Isa,~L.; Dufresne,~E.~R. Adsorption of soft particles at fluid
	interfaces. \emph{Soft Matter} \textbf{2015}, \emph{11}, 7412--7419
	\bibitem[Mehrabian \latin{et~al.}(2016)Mehrabian, Harting, and
	Snoeijer]{2016Mehrabin}
	Mehrabian,~H.; Harting,~J.; Snoeijer,~J.~H. Soft particles at a fluid
	interface. \emph{Soft Matter} \textbf{2016}, \emph{12}, 1062--1073
	\bibitem[Geisel \latin{et~al.}(2012)Geisel, Isa, and Richtering]{2012Geisel}
	Geisel,~K.; Isa,~L.; Richtering,~W. Unraveling the 3D localization and
	deformation of responsive microgels at oil/water interfaces: a step forward
	in understanding soft emulsion stabilizers. \emph{Langmuir} \textbf{2012},
	\emph{28}, 15770--15776
	\bibitem[Geisel \latin{et~al.}(2015)Geisel, Henzler, Guttmann, and
	Richtering]{2015GeiselTXM}
	Geisel,~K.; Henzler,~K.; Guttmann,~P.; Richtering,~W. New Insight into
	Microgel-Stabilized Emulsions Using Transmission X-ray Microscopy: Nonuniform
	Deformation and Arrangement of Microgels at Liquid Interfaces.
	\emph{Langmuir} \textbf{2015}, \emph{31}, 83--89
	\bibitem[hin Kwok and Ngai(2016)hin Kwok, and Ngai]{2016Kwok}
	hin Kwok,~M.; Ngai,~T. A confocal microscopy study of micron-sized
	poly(N-isopropylacrylamide) microgel particles at the oil-water interface and
	anisotopic flattening of highly swollen microgel. \emph{J. Colloid Interface
		Sci.} \textbf{2016}, \emph{461}, 409 -- 418
	\bibitem[Richtering(2012)]{2012Richtering}
	Richtering,~W. Responsive Emulsions Stabilized by Stimuli-Sensitive Microgels:
	Emulsions with Special Non-Pickering Properties. \emph{Langmuir}
	\textbf{2012}, \emph{28}, 17218--17229
	\bibitem[Rey \latin{et~al.}(2016)Rey, Elnathan, Ditcovski, Geisel, Zanini,
	Fernandez-Rodriguez, Naik, Frutiger, Richtering, Ellenbogen, Voelcker, and
	Isa]{2016Rey}
	Rey,~B.~M.; Elnathan,~R.; Ditcovski,~R.; Geisel,~K.; Zanini,~M.;
	Fernandez-Rodriguez,~M.-A.; Naik,~V.~V.; Frutiger,~A.; Richtering,~W.;
	Ellenbogen,~T.; Voelcker,~N.~H.; Isa,~L. Fully tunable silicon nanowire
	arrays fabricated by soft nanoparticle templating. \emph{Nano Lett.}
	\textbf{2016}, \emph{16}, 157--163
	\bibitem[Scheidegger \latin{et~al.}(2017)Scheidegger, Fernandez-Rodriguez,
	Geisel, Zanini, Elnathan, Richtering, and Isa]{2017Scheidegger}
	Scheidegger,~L.; Fernandez-Rodriguez,~M.~A.; Geisel,~K.; Zanini,~M.;
	Elnathan,~R.; Richtering,~W.; Isa,~L. Compression and deposition of microgel
	monolayers from fluid interfaces: particle size effects on interface
	microstructure and nanolithography. \emph{Phys. Chem. Chem. Phys.}
	\textbf{2017}, \emph{19}, 8671--8680
	\bibitem[D{\"o}rr \latin{et~al.}(2016)D{\"o}rr, Hardt, Masoud, and
	Stone]{2016Dorr}
	D{\"o}rr,~A.; Hardt,~S.; Masoud,~H.; Stone,~H.~A. Drag and diffusion
	coefficients of a spherical particle attached to a fluid-fluid interface.
	\emph{J. Fluid Mech.} \textbf{2016}, \emph{790}, 607–618
	\bibitem[Danov \latin{et~al.}(1995)Danov, Aust, Durst, and Lange]{1995Danov}
	Danov,~K.; Aust,~R.; Durst,~F.; Lange,~U. Influence of the surface viscosity on
	the hydrodynamic resistance and surface diffusivity of a large Brownian
	particle. \emph{J. Colloid Interf. Sci.} \textbf{1995}, \emph{175}, 36 --
	45
	\bibitem[Fischer \latin{et~al.}(2006)Fischer, Dhar, and Heinig]{2006Fischer}
	Fischer,~T.~M.; Dhar,~P.; Heinig,~P. The viscous drag of spheres and filaments
	moving in membranes or monolayers. \emph{J. Fluid Mech.} \textbf{2006},
	\emph{558}, 451–475
	\bibitem[Taylor(1982)]{1982Taylor}
	Taylor,~J.~R. \emph{An Introduction to Error Analysis: The Study of
		Uncertainties in Physical Measurements}; University Science Books, Sausalito,
	California, 1982
	\bibitem[Cerbino and Trappe(2008)Cerbino, and Trappe]{2008Cerbino}
	Cerbino,~R.; Trappe,~V. Differential dynamic microscopy: Probing wave vector
	dependent dynamics with a microscope. \emph{Phys. Rev. Lett.} \textbf{2008},
	\emph{100}, 188102
	\EndOfBibitem
	\bibitem[Bayles \latin{et~al.}(2016)Bayles, Squires, and Helgeson]{2016Bayles}
	Bayles,~A.~V.; Squires,~T.~M.; Helgeson,~M.~E. Dark-field differential dynamic
	microscopy. \emph{Soft Matter} \textbf{2016}, \emph{12}, 2440--2452
	\bibitem[Crocker and Grier(1996)Crocker, and Grier]{1996Crocker}
	Crocker,~J.~C.; Grier,~D.~G. Methods of digital video microscopy for colloidal
	studies. \emph{J. Colloid Interf. Sci.} \textbf{1996}, \emph{179}, 298 --
	310
	\bibitem[Isa \latin{et~al.}(2011)Isa, Amstad, Schwenke, Del~Gado, Ilg,
	Kr\"{o}ger, and Reimhult]{2011Isa2}
	Isa,~L.; Amstad,~E.; Schwenke,~K.; Del~Gado,~E.; Ilg,~P.; Kr\"{o}ger,~M.;
	Reimhult,~E. Adsorption of core-shell nanoparticles at liquid-liquid
	interfaces. \emph{Soft Matter} \textbf{2011}, \emph{7}, 7663--7675
	\bibitem[Zell \latin{et~al.}(2014)Zell, Isa, Ilg, Leal, and Squires]{2014Zell}
	Zell,~Z.~A.; Isa,~L.; Ilg,~P.; Leal,~L.~G.; Squires,~T.~M. Adsorption energies
	of poly(ethylene oxide)-based surfactants and nanoparticles on an air-water
	surface. \emph{Langmuir} \textbf{2014}, \emph{30}, 110--119
	\bibitem[Halperin and Kr\"oger(2011)Halperin, and Kr\"oger]{Halperin2011}
	Halperin,~A.; Kr\"oger,~M. Collapse of thermoresponsive brushes and the tuning
	of protein adsorption. \emph{Macromolecules} \textbf{2011}, \emph{44},
	6986--7005
	\bibitem[Doi(1996)]{1996Doi}
	Doi,~M. \emph{Introduction to Polymer Physics}; Oxford University Press,
	1996
	\bibitem[Lopez and Richtering(2017)Lopez, and Richtering]{2017Lopez}
	Lopez,~C.~G.; Richtering,~W. Does Flory-Rehner theory quantitatively describe
	the swelling of thermoresponsive microgels? \emph{Soft Matter} \textbf{2017},
	\emph{13}, 8271--8280
	\bibitem[Fern\'andez-Barbero \latin{et~al.}(2002)Fern\'andez-Barbero,
	Fern\'andez-Nieves, Grillo, and L\'opez-Cabarcos]{2002Barbero}
	Fern\'andez-Barbero,~A.; Fern\'andez-Nieves,~A.; Grillo,~I.;
	L\'opez-Cabarcos,~E. Structural modifications in the swelling of
	inhomogeneous microgels by light and neutron scattering. \emph{Phys. Rev. E}
	\textbf{2002}, \emph{66}, 051803
	\bibitem[Lang \latin{et~al.}(2017)Lang, Lenart, Sun, Hammouda, and
	Hore]{2017Lang}
	Lang,~X.; Lenart,~W.~R.; Sun,~J. E.~P.; Hammouda,~B.; Hore,~M. J.~A.
	Interaction and Conformation of Aqueous Poly(N-isopropylacrylamide) (PNIPAM)
	Star Polymers below the LCST. \emph{Macromolecules} \textbf{2017}, \emph{50},
	2145--2154
	\bibitem[Stieger \latin{et~al.}(2004)Stieger, Richtering, Pedersen, and
	Lindner]{2004Stieger}
	Stieger,~M.; Richtering,~W.; Pedersen,~J.~S.; Lindner,~P. Small-angle neutron
	scattering study of structural changes in temperature sensitive microgel
	colloids. \emph{J. Chem. Phys.} \textbf{2004}, \emph{120}, 6197--6206
	\bibitem[Halperin \latin{et~al.}(2015)Halperin, Kr\"oger, and
	Winnik]{Halperin2015}
	Halperin,~A.; Kr\"oger,~M.; Winnik,~F.~M. Poly(N-isopropylacrylamide) phase
	diagrams: Fifty years of research. \emph{Angew. Chem. Int. Ed.}
	\textbf{2015}, \emph{54}, 15342--15367
	\bibitem[Huang \latin{et~al.}(2016)Huang, Gawlitza, von Klitzing, Gilson,
	Nowak, Odenbach, Steffen, and Auernhammer]{2016Huang}
	Huang,~S.; Gawlitza,~K.; von Klitzing,~R.; Gilson,~L.; Nowak,~J.; Odenbach,~S.;
	Steffen,~W.; Auernhammer,~G.~K. Microgels at the water/oil interface: In situ
	observation of structural aging and two-dimensional magnetic bead
	microrheology. \emph{Langmuir} \textbf{2016}, \emph{32}, 712--722
	\bibitem[Vasudevan \latin{et~al.}(2018)Vasudevan, Rauh, Barbera, Karg, and
	Isa]{2018Vasudevan}
	Vasudevan,~S.~A.; Rauh,~A.; Barbera,~L.; Karg,~M.; Isa,~L. Stable in bulk and
	aggregating at the interface: Comparing core–shell nanoparticles in
	suspension and at fluid interfaces. \emph{Langmuir} \textbf{2018}, \emph{34},
	886--895
	\bibitem[Routh and Zimmerman(2003)Routh, and Zimmerman]{2003Routh}
	Routh,~A.~F.; Zimmerman,~W.~B. The diffusion coefficient of a swollen microgel
	particle. \emph{J. Colloid Interf. Sci.} \textbf{2003}, \emph{261}, 547 --
	551
	\bibitem[Isa \latin{et~al.}(2011)Isa, Lucas, Wepf, and Reimhult]{2011Isa1}
	Isa,~L.; Lucas,~F.; Wepf,~R.; Reimhult,~E. Measuring single-nanoparticle
	wetting properties by freeze-fracture shadow-casting cryo-scanning electron
	microscopy. \emph{Nat. Commun.} \textbf{2011}, \emph{2}, 438
	\bibitem[\'{A}lvarez Puebla \latin{et~al.}(2009)\'{A}lvarez Puebla,
	Contreras-C\'{a}ceres, Pastoriza-Santos, P\'{e}rez-Juste, and
	Liz-Marz\'{a}n]{2009Alvarez}
	\'{A}lvarez Puebla,~R.; Contreras-C\'{a}ceres,~R.; Pastoriza-Santos,~I.;
	P\'{e}rez-Juste,~J.; Liz-Marz\'{a}n,~L. Au@pNIPAM colloids as molecular traps
	for surface-enhanced, spectroscopic, ultra-sensitive analysis. \emph{Angew.
		Chem. Int. Ed.} \textbf{2009}, \emph{48}, 138--143
	\bibitem[Yang \latin{et~al.}(2016)Yang, Hu, Keiper, Xiong, and
	Hallinan]{2016Yang}
	Yang,~G.; Hu,~L.; Keiper,~T.~D.; Xiong,~P.; Hallinan,~D.~T. Gold nanoparticle
	monolayers with tunable optical and electrical properties. \emph{Langmuir}
	\textbf{2016}, \emph{32}, 4022--4033
	\bibitem[Karg \latin{et~al.}(2011)Karg, Jaber, Hellweg, and Mulvaney]{2011Karg}
	Karg,~M.; Jaber,~S.; Hellweg,~T.; Mulvaney,~P. Surface plasmon spectroscopy of
	gold-poly-N-isopropylacrylamide core-shell particles. \emph{Langmuir}
	\textbf{2011}, \emph{27}, 820--827
	\bibitem[Boniello \latin{et~al.}(2015)Boniello, Blanc, Fedorenko, Medfai,
	Mbarek, In, Gross, Stocco, and Nobili]{2015Boniello}
	Boniello,~G.; Blanc,~C.; Fedorenko,~D.; Medfai,~M.; Mbarek,~N.~B.; In,~M.;
	Gross,~M.; Stocco,~A.; Nobili,~M. Brownian diffusion of a partially wetted
	colloid. \emph{Nat. Mater.} \textbf{2015}, \emph{14}, 908
\end{thebibliography}

\begin{tocentry}
	\centering
	\includegraphics[scale = 0.85]{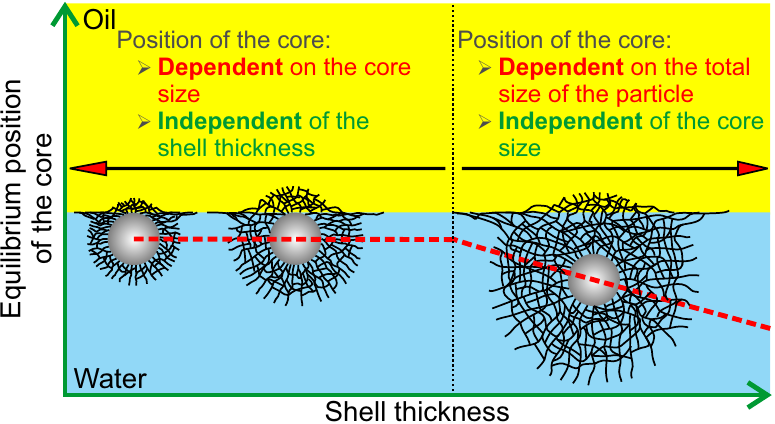}
\end{tocentry}

\end{document}